\documentclass[arxiv]{frontiersSCNS} 

\usepackage{url,hyperref,lineno,microtype,subcaption}
\hypersetup{
    colorlinks=true,
    linkcolor=cyan,
    filecolor=gray,      
    urlcolor=blue,
    citecolor=darkgray
}
\usepackage[onehalfspacing]{setspace}
\usepackage[T1]{fontenc}    
\usepackage{booktabs}       
\usepackage{amsfonts}       
\usepackage{nicefrac}       
\usepackage{lipsum}
\usepackage{todonotes}
\usepackage{amsmath}
\usepackage{bm}
\usepackage{slashed}
\usepackage{xspace}
\newcommand{\pt}{\ensuremath{p_{\mathrm{T}}}\xspace}
\newcommand{\GeV}{\ensuremath{\,\text{Ge\hspace{-.08em}V}}\xspace}
\newcommand{\TeV}{\ensuremath{\,\text{Te\hspace{-.08em}V}}\xspace}
\newcommand{\ptmiss}{\ensuremath{\pt^\text{miss}}\xspace}
\newcommand{\ptvecmiss}{\ensuremath{{\vec p}_{\mathrm{T}}^{\kern1pt\text{miss}}}\xspace}
\newcommand{\HT}{\ensuremath{H_{\mathrm{T}}}\xspace}
\newcommand{\PW}{\ensuremath{\mathrm{W}}\xspace}

\newcommand{\PZ}{\ensuremath{\mathrm{Z}}\xspace}
\newcommand{\PZpr}{\ensuremath{\PZ^\prime}\xspace}
\newcommand{\PQb}{\ensuremath{\mathrm{b}}\xspace}
\newcommand{\PQt}{\ensuremath{\mathrm{t}}\xspace}

\newcommand{\Pgm}{\ensuremath{\mu}\xspace}
\newcommand{\Pe}{\ensuremath{\mathrm{e}}\xspace}
\newcommand{\PSgxpm}{\ensuremath{{\widetilde{\chi}^\pm}}}
\newcommand{\PSgxz}{\ensuremath{{\widetilde{\chi}^0}}}
\newcommand{\PSg}{\ensuremath{\mathrm{\widetilde{g}}}}
\newcommand{\PSq}{\ensuremath{\mathrm{\widetilde{q}}}}
\newcommand{\PSt}{\ensuremath{\mathrm{\widetilde{t}}}}
\ifarxiv
\else
\linenumbers
\fi

\def\firstAuthorLast{Jawahar {et~al.}}
\def\Authors{Pratik~Jawahar\,$^{1,\ast}$, Thea~Aarrestad\,$^{1}$,
Nadezda Chernyavskaya$^{1}$,
Maurizio~Pierini\,$^{1}$, Kinga~A.~Wozniak\,$^{1,2}$, Jennifer~Ngadiuba\,$^{3,4}$, Javier~Duarte\,$^{5}$,
Steven~Tsan\,$^{5}$}
\def\Address{$^{1}$European Center for Nuclear Research (CERN),
   CH 1211, Geneva 23, Switzerland\\
$^{2}$University of Vienna, 1010 Wien, Austria\\
$^{3}$Fermi National Accelerator Laboratory (FNAL), Batavia, IL 60510, USA\\
$^{4}$California Institute of Technology, Pasadena, CA 91125, USA\\
$^{5}$University of California San Diego, La Jolla, CA 92093, USA}
\def\corrAuthor{Pratik~Jawahar}
\def\corrEmail{pjawahar@wpi.edu}

\begin{document}
\onecolumn
\firstpage{1}

\title{Improving Variational Autoencoders for New Physics Detection at the LHC with Normalizing Flows}

\author[\firstAuthorLast ]{\Authors}
\address{} 
\correspondance{} 

\extraAuth{}

\maketitle
\begin{abstract}
We investigate how to improve new physics detection strategies exploiting variational autoencoders and normalizing flows for anomaly detection at the Large Hadron Collider. 
As a working example, we consider the DarkMachines challenge dataset. 
We show how different design choices (e.g., event representations, anomaly score definitions, network architectures) affect the result on specific benchmark new physics models. 
Once a baseline is established, we discuss how to improve the anomaly detection accuracy by exploiting normalizing flow layers in the latent space of the variational autoencoder.
\end{abstract}



\section{Introduction}

Most searches for new physics at the CERN Large Hadron Collider (LHC) target specific experimental signatures. 
The underlying assumption of a specific new physics model could enter at various stages in the search design, e.g., when reducing the data rate from 40\,M to 1,000 collision events per second in real time~\citep{ATLASL1T,CMSL1T,CMSHLT}, when designing the event selection, or when running the final hypothesis testing. 
When searching for pre-established and theoretically well-motivated particles (e.g., the Higgs boson), this strategy is extremely successful because the underlying assumption can be exploited to maximize the search sensitivity. 
On the other hand, the lack of a predefined target might turn this strength into a limitation. 

To compensate for this potential problem, \emph{model independent} searches are also carried out~\citep{Aaboud:2018ufy,Aaltonen:2008vt,Aaron:2008aa,CMS-PAS-EXO-14-016,Abazov:2011ma} at hadron colliders. 
These searches consist in an extensive set of comparisons between the data distribution and the expectation derived from Monte Carlo simulation. 
Many comparisons are carried out in parallel for multiple physics-motivated features while applying different event selections. 
However, when searching for new physics among many channels, the ``global'' significance of observing a particular discrepancy must take into account the probability of observing such a discrepancy anywhere. 
This so called look-elsewhere effect can be quantified in terms of a trial factor~\citep{Gross:2010qma}.
While the large trial factor typically reduces the statistical power of this strategy in terms of significance, model independent searches are valuable tools to identify possible regions of interest and provide data-driven motivations for traditional, more targeted searches to be performed on future data.

Recently, the use of machine learning techniques has been advocated as a mean to reduce the model dependence~\citep{Weisser:2016cnc,Cerri:2018anq,DAgnolo:2018cun,DeSimone:2018efk,Farina:2018fyg,Collins:2018epr,Blance:2019ibf,Hajer:2018kqm,Heimel:2018mkt,Collins:2019jip,DAgnolo:2019vbw,Nachman:2020lpy,Andreassen_2020,Amram:2020ykb,Dillon:2020quc,Cheng:2020dal,Khosa:2020qrz,Nachman:2020ccu,Park:2020pak,Bortolato:2021zic,Collins:2021nxn,Finke:2021sdf,Gonski:2021jek,Hallin:2021wme,ostdiek2021deep}. 
In this context, the particle-physics community engaged in two data challenges: the LHC Olympics 2020~\citep{kasieczka2021lhc} and the DarkMachines challenge~\citep{aarrestad2021dark}, where different approaches were explored to attempt to detect an unknown signal of new physics hidden in simulated data.

As part of our contribution to the DarkMachines challenge, we investigated the use of a particle-based variational autoencoder (VAE)~\citep{kingma2013auto,pmlr-v32-rezende14} and the possibility of enhancing its anomaly detection capability by using normalizing flows (NFs)~\citep{papamakarios2021normalizing} in the latent space to optimize the choice of the latent-space prior. 
In this paper, we document those studies and expand that effort, investigating the impact of specific architecture choices (event representation, network architecture, usage of expert features, and definition of the anomaly score). 
This study is an update of our contribution to  the DarkMachine challenge~\citep{aarrestad2021dark}, which benefits from the lessons learned by the DarkMachines challenge. 
Taking inspiration from solutions presented by other groups in the challenge (e.g., Refs.~\citep{Caron:2021wmq,ostdiek2021deep}), we evaluate the impact of some of their findings on our specific setup. 
In some cases (but not always), these solutions translate in an improved performance, quantified using the same metrics presented in \cite{aarrestad2021dark}. 
In this way, we establish an improved baseline model, on top of which we evaluate the impact of the normalizing flow layers in the latent space. 

\section{Data samples and event representation}
\label{sec:dataset_desc}
This study is based on the datasets released on the Zenodo platform~\citep{darkmachines_community_2020_3961917} in relation to the Dark Machines Anomaly Score Challenge~\citep{aarrestad2021dark}. 
They consist of a set of processes predicted in the standard model (SM) of particle physics, mixed according to their production cross section in proton-proton collisions at 13\TeV center-of-mass energy, and a set of benchmark signal samples. 
The datasets contains labels, identifying the process that generated each event. 
Labels are ignored during training and used to evaluate performance metrics.

For each sample, four datasets are provided, corresponding to four 
different event selections (called \emph{channels}~\citep{aarrestad2021dark}):
\begin{itemize}
    \item Channel 1: $\HT \geq 600\GeV$, $\ptmiss\geq 200\GeV$, and $\ptmiss/\HT \geq 0.2$. 
    \item Channel 2a: $\ptmiss\geq 50\GeV$ and at least three light leptons (muons or electrons) with  $\pt>15\GeV$.
    \item Channel 2b: $\ptmiss\geq 50\GeV$, $\HT\geq 50\GeV$ and at least two light leptons (muons or electrons) with  $\pt>15\GeV$.
    \item Channel 3:  $\HT \geq 600\GeV$, $\ptmiss\geq 100\GeV$.
\end{itemize} 
where $\pt$ is the magnitude of a particle's transverse momentum, $\HT$ is the scalar sum of the jet $\pt$ in the event, and $\ptvecmiss$ is the vector equal and opposite to the vector sum of the transverse momenta of the reconstructed particles in the event, while $\ptmiss$ is its magnitude\footnote{We use a Cartesian coordinate system with the $z$ axis oriented along the beam axis, the $x$ axis on the horizontal plane, and the $y$ axis oriented upward. 
The $x$ and $y$ axes define the transverse plane, while the $z$ axis identifies the longitudinal direction. 
The azimuth angle $\phi$ is computed with respect to the $x$ axis. 
The polar angle $\theta$ is used to compute the pseudorapidity $\eta = -\log(\tan(\theta/2))$. 
The transverse momentum ($\pt$) is the projection of the particle momentum on the ($x$, $y$) plane. 
We fix units such that $c=\hbar=1$.}.
More details are provided in \cite{aarrestad2021dark}. 

\begin{table}[ht!]
    \centering
    \caption{Summary of the available dataset size.\label{tab:dataset_size}}
    \begin{tabular}{c|cccc}
        Dataset &  Channel 1 & Channel 2a & Channel 2b & Channel 3 \\
        \hline
        Training  & $193,800$ & $13,425$ & $238,450$ & $7,100,934$ \\
        Validation & $10,200$  & $707$  & $12,550$ & $373,733$ \\
        Bkg. Test & $10,000$ & $5,868$ & $89,000$ & $1,025,333$\\
        Sig. Test & $38,666$ & $5,868$ & $89,676$ & $1,023,320$
    \end{tabular}
    \end{table}
    
The input consists of the momenta of all the reconstructed physics objects in the event (jets, \PQb jets, electrons \Pe, muons \Pgm, and photons), ordered by decreasing $\pt$. 
Each list of objects is zero-padded to force each event into a fixed-length matrix with the same order: up to 15 jets, and up to 4 each of \PQb jets, $\Pgm^{\pm}$, $\Pe^{\pm}$, and photons. 
We pre-process the input by applying the \texttt{scikit-learn}~\citep{scikit-learn} standard scaling and arranging the list of objects into a matrix of 39 particles times four momentum features $(E, \pt, \eta, \phi)$, where $E$ is the particle energy. For \Pe, \Pgm, and photons, the energy is computed assuming zero mass. For jets, the measured jet mass is used.
The input matrix is interpreted as an image or an unordered point cloud, depending on the underlying VAE architecture.

The training and validation dataset consists of background events from the SM mixture. 
The available dataset size is detailed in Table~\ref{tab:dataset_size} for each of the channels. 
The background test samples are combined with the benchmark signal samples listed in Table~\ref{tab:sig_models} to form the labeled test dataset on which performance is evaluated.

\begin{table}[ht!]
    \centering
    \caption{BSM processes contributing to the signal dataset in each channel. The process code, adopted in this study, is taken from \cite{aarrestad2021dark}. \label{tab:sig_models}}
    \begin{tabular}{c|c|c|c|c|c}
        BSM process & Code & Ch.1 & Ch.2a & Ch.2b & Ch.3\\
        \hline
        \PZpr + jet & \texttt{monojet\_Zp2000.0\_DM\_50.0} & $\times$
         & $\times$  & & $\times$\\
        \PZpr+ \PW/\PZ & \texttt{monoV\_Zp2000.0\_DM\_50.0} & 
         &  & & $\times$\\
        \PZpr + \PQt & \texttt{monotop\_200\_A} & $\times$ & & & $\times$\\
        \PZpr in LFV $\mathrm{U}(1)_{\mathrm{L}_{\mu}-\mathrm{L}_{\tau}}$ & \texttt{pp23mt\_50} &  & $\times$ & $\times$ & \\
        & \texttt{pp24mt\_50} &  & $\times$ & $\times$ & \\
        $\slashed{R}$-SUSY $\PSt\PSt$ & \texttt{stlp\_st1000} & $\times$ & & $\times$ &$\times$ \\
        $\slashed{R}$-SUSY $\PSq\PSq$ & \texttt{sqsq1\_sq1400\_neut800} & $\times$ & & &$\times$ \\
        SUSY $\PSg\PSg$ & \texttt{glgl1400\_neutralino1100} & $\times$ & $\times$ & $\times$ & $\times$\\
        & \texttt{glgl1600\_neutralino800} & $\times$ & $\times$ & $\times$ & $\times$\\
        SUSY $\PSt\PSt$ & \texttt{stop2b1000\_neutralino300} & $\times$ & & & $\times$ \\
        SUSY $\PSq\PSq$ & \texttt{sqsq\_sq1800\_neut800} & $\times$ & & & $\times$\\
        SUSY $\PSgxpm\PSgxz$ & \texttt{chaneut\_cha200\_neut50} &  & $\times$ & $\times$ &\\
        & \texttt{chaneut\_cha250\_neut150} &  & $\times$ & $\times$ &\\
        SUSY $\PSgxpm\PSgxpm$ & \texttt{chacha\_cha300\_neut140} &  &  & $\times$ &\\
        & \texttt{chacha\_cha400\_neut60} &  &  & $\times$ &\\
        & \texttt{chacha\_cha600\_neut200} &  &  & $\times$ &
\end{tabular}
\label{table:bsm_procs}
\end{table}

\section{Training setup and evaluation metrics}
\label{sec:training_setup_metrics}

Variational Autoencoders~\citep{kingma2013auto,pmlr-v32-rezende14,kingma2019introduction} are a class of likelihood-based generative models that maximize the likelihood of the training data according to the generative model $\prod_{x \in \mathrm{data}}p_\theta(x)$ for the set of observed variables $x$ in the training data. 
To achieve this in a tractable way, the generative model is augmented by the introduction of a set of latent variables $z$, such that the the marginal distribution over the observed variables $p_\theta(x)$, is given
by: $p_\theta(x) = \int p_\theta(x | z)q_\theta(z)dz$. 
In this way, $q_\theta(z)$ can be a relatively simple distribution, such as a Gaussian, while maintaining high expressivity for the marginal distribution $p_\theta(x)$ as an infinite mixture of simple distributions controlled by $z$. 
Besides being used as generative models, VAEs have been shown to be effective as anomaly detection algorithms~\citep{an2015variational}.

In this work, the VAE models are trained on the training and validation datasets, minimizing the loss function:
\begin{equation}
    \label{eq:VAELoss}
    L_\mathrm{total} = \beta D_\mathrm{KL} + (1-\beta) L_\mathrm{C}~,
\end{equation}
where $L_\mathrm{C}$ is a reconstruction loss, which is chosen to be an L$_1$-type permutation-invariant Chamfer loss~\citep{10.5555/1622943.1622971}:
\begin{equation}
    \label{eq:ChamferLoss}
    L_\mathrm{C} = \sum_{\vec x \in S_\mathrm{input}} \underset{\vec y \in S_\mathrm{output}}{\mathrm{min}} \big|\vec{x} - \vec{y}\big|
    +
    \sum_{\vec y \in S_\mathrm{output}} \underset{\vec x \in S_\mathrm{input}}{\mathrm{min}} \big|\vec{x} - \vec{y}\big|~,
\end{equation}
similar to the L$_2$-type Chamfer distance used in Refs.~\citep{Fan_2017_CVPR,Zhang2020FSPool}.
In Eq.~(\ref{eq:ChamferLoss}), $D_\mathrm{KL}$ is the Kullback--Liebler divergence term usually employed to force the data distribution in the latent space to a multidimensional Gaussian with unitary covariance matrix~\citep{rezende2015variational}, and $\beta$ is a parameter that controls the relative importance of the two terms~\citep{Higgins2017betaVAELB}. 

All of our models are optimized using the Adam minimizer~\citep{kingma2017adam}. 
A learning rate of $10^{-4}$ is applied along with a brute force early stopping strategy used on an ad-hoc basis. 
A batch size of 32 is chosen to train models. 
All models are implemented with the \texttt{PyTorch}~\citep{NEURIPS2019_9015} deep learning framework and are hosted on GitHub~\citep{Pratik_code}.

We train and test all our models on the WPI Turing Research Cluster\footnote{\url{https://arc.wpi.edu/computing/hpc-clusters/}}, using 8 CPU nodes and 1 GPU node (NVIDIA Tesla V100 or Tesla P100).

At inference time, $L_\mathrm{C}$ is used as an anomaly detection score, to quantify the distance between the input and the output. 
By applying a lower-bound threshold on $L_\mathrm{C}$, we identify every event with an $L_\mathrm{C}$ value larger than the threshold as an anomaly. 
By comparing this prediction to the ground truth, we can assess the performance of the VAE on specific signal benchmark models. 

To evaluate model performance we follow the same strategy and code used in \cite{aarrestad2021dark} to enable comparison with other models tested on this dataset. As explained in \cite{aarrestad2021dark}, we extract four main performance parameters from the receiver operating characteristic (ROC) curves based on the chosen anomaly metric for each model, namely the area under the curve (AUC) and true positive rate (also known as the signal efficiency $\epsilon_\mathrm{S}$) at three different, fixed values of the false positive rate (also known as background efficiency $\epsilon_\mathrm{B}$). 
We then combine these scores from all models on all available signal regions across all channels of the dataset to form box-and-whisker plots, using 6 different combination and comparison strategies namely, the highest mean score method, highest median score method, average rank method, top scorer method, top-5 scorer method, and highest minimum scorer method. 
A box is drawn spanning the inner half (50\% quantile centered at the median) of the data as shown in Fig.~\ref{fig:input_encodings}. A line through the box marks the median. Whiskers extend from the box to either the maximum and minimum unless these are further away from the edge of the box than 1.5 box lengths. 
The outlier points are shown as circles.

For Fig.~\ref{fig:input_encodings} and the other figures, the representative ranking as denoted by the legend corresponds to the performance based on the highest mean score method unless mentioned otherwise.
However, to choose the best model for each experiment described in this paper, we consider all six comparison methods to arrive at a consensus. 
The code to perform these comparisons and to generate the corresponding plots is available in \cite{aarrestad2021dark}.

\section{Baseline VAE model}
\label{sec:convVAE}

The main goal of this study is to evaluate the impact of normalizing flow layers in the latent space on the anomaly detection capability of a reference VAE model. 
This and the following sections describe how this reference model is built, starting from the VAE based on convolutional layers (Conv-VAE) presented in \cite{aarrestad2021dark} and modifying its architecture based on some of the lessons learned during the DarkMachine challenge.

The encoder of the initial Conv-VAE consists of three convolutional layers, with 32, 16, and 8 kernels of size $(3,4)$, $(5,1)$, and $(7,1)$, respectively. 
For all layers, the stride is set to 1 and zero padding to ``same''.
The output of the convolutional layers is flattened and passed to 2 fully-connected neural network (FCN) layers that output the mean and variance for the latent space. 
The cardinality of the latent space is fixed to 15. 
The decoder mirrors the encoder architecture, returning an output of the same size as the input.

In order to define the reference model, the architecture of the starting model is modified in different ways, each time evaluating the impact of a given choice on the test dataset. 
Several possibilities are considered: how to embed the event in the two-dimensional ($2$D) array (see section~\ref{sec:data_representation}); how to interpret the array, e.g., as an image or a graph (see section~\ref{sec:image_or_graph}); whether to extend the event representation beyond the particle momenta, adding domain-specific high level features as an additional input (see section~\ref{sec:event_variables}); and which anomaly score to use (see section~\ref{sec:anomaly score}). 
We study various options for each of these points, following this order. 
Doing so, we establish a candidate model, which replaces the initial model. 
We evaluate on this new model the benefit of using normalizing flow layers in the latent space (see section~\ref{sec:VAE_Flows}) to improve the anomaly detection accuracy.\\

\subsection{Data representation}
\label{sec:data_representation}

\begin{figure}[t!]
    \centering
    \includegraphics[width=0.8\linewidth]{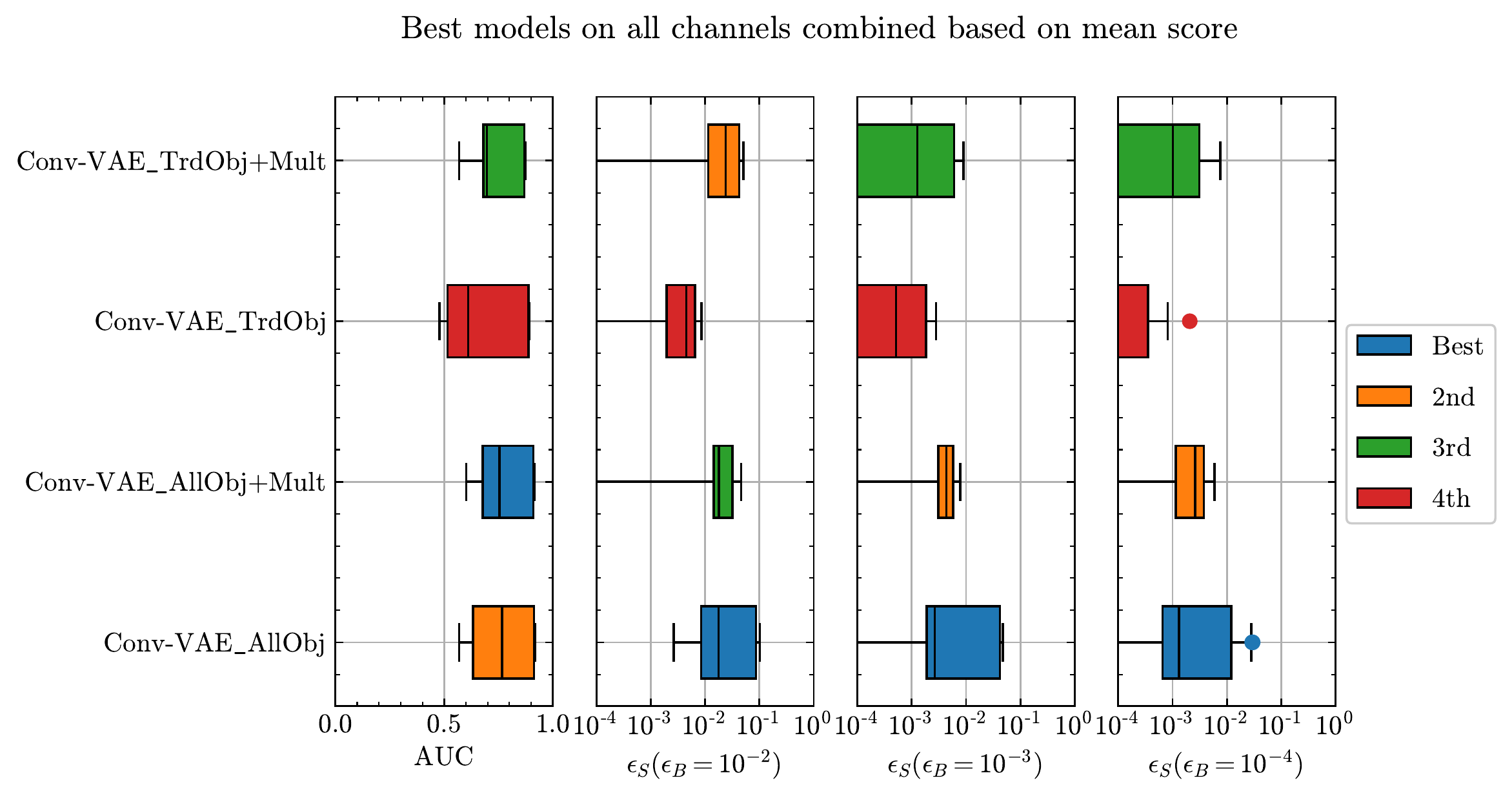}
    \caption{Anomaly detection performance for the Conv-VAE with different inputs given (see text for more details): all physics objects in the event (AllObj); truncated input object list (TrdObj); all objects and array of object multiplicity (AllObj+Mult); truncated input object list and array of object multiplicity (TrdObj+Mult).
    \label{fig:input_encodings}}
\end{figure}

By their nature, events consist of a variable number of objects. 
To some extent, this conflicts with most neural network architectures, which assume a fixed-size input. 
As a baseline, we adopt the simplest solution, i.e., to zero-pad all events to standardized event sizes for all available samples. 
To get a better idea of how padding affects results, we study performance across alternative input encodings. 
We consider two main types of encodings, listed as AllObj and TrdObj in Fig.~\ref{fig:input_encodings}. 
The former involves considering the entire event which implies allowing for a large enough padding such that every object per event is taken into consideration across the entire dataset. 
The latter involves cutting down the padding and the input sequence by considering only up to four leading jets and three objects each of the other types per event. 

When using the truncated sequence, the model loses information regarding the number of objects of each type per event, which is implicitly learned when the whole sequence is considered. 
To compensate for this loss, one can explicitly add this information passing a second input to the model, consisting of a vector containing the multiplicities of each object type. 
This input is concatenated to the flattened output received from the convolutional layers in the encoder before passing them to the fully connected layers. 
For the sake of comparison, we also do the same for the AllObj case (labeled as ``+Mult'' in Fig.~\ref{fig:input_encodings}).

The results in Fig.~\ref{fig:input_encodings} 
show that the truncated sequence does worse than the full sequence. 
We also see little improvement in performance with the addition of multiplicity information per event in both the AUC as well as performance at lower background efficiencies. 
As a result, we keep the input encoding that considers the complete sequence per event.\\ 

\subsection{VAE architecture}
\label{sec:image_or_graph}

The convolutional architecture used for the baseline VAE is not the only option to handle the input considered in this study. 
The ensemble of reconstructed particles in an event can be represented as a point cloud. 
Doing so, we can process it with a graph neural network. 
The main advantage of this choice stands with the permutation invariance of the graph processing, which pairs that of the loss in Eq.~\ref{eq:ChamferLoss} and complies with the unordered nature of the input list of particles. 
Graph-based architectures have already been shown to perform better with sparse, non-Euclidean data representations in general~\citep{7974879, zhou2020graph} and in particle physics in particular~\citep{Shlomi:2020gdn,duarte2020graph}.

To this end, we consider a GCN-VAE model composed of multilayer graph convolutional network layers (GCNs)~\citep{kipf2016semi} and FCN layers in both the encoder and the decoder. 
As for the VAE, the input graphs are built from the input list described in section~\ref{sec:dataset_desc}, each particle representing one vertex of the graph in the space identified by five particle features: $E,\pt,\eta,\phi$, and object type. 
The object type is a label-encoded integer that signifies the object type. 
The input is structured as a fully connected, undirected graph which is passed to the GCN layers of the encoder, defined as~\citep{kipf2016semi}:
\begin{equation}
    H_{(l+1)} = \sigma(\widetilde{D}^{-\frac{1}{2}}\widetilde{A}\widetilde{D}^{-\frac{1}{2}}H_{(l)}W_{(l)})~,
\end{equation}
where $H_{(l)}$ is the input to the $(l+1)$th GCN layer with $H_{(0)} = X$ where $X$ represents the node feature matrix. 
$H_{(l+1)}$ is the layer output, $\widetilde{A}= A + I$, where $A$ is the adjacency of the graph, with $I$ being the identity matrix which implies added self connections for each node. 
$\widetilde{D}_{ii}=\sum_{j}A_{ij}$ is defined for the normalized adjacency based message passing regime, $W_{(l)}$ is the layer weights matrix and $\sigma(\bullet)$ is a suitable nonlinear activation function. 
The output of the last GCN layer is flattened and passed to an FCN layer which populates the latent space. 
The encoder has 3 GCN layers that scale the 5 node features to 32, 16, and 2 respectively, followed by a single FCN layer which generates a 15-dimensional latent space. 
The decoder has a symmetrically inverted structure with the sampled point being upscaled through an FCN layer first and the resulting output is reshaped and passed to GCN layers that reconstruct the node features.

\begin{figure}[t!]
    \centering
    \includegraphics[width=0.8\linewidth]{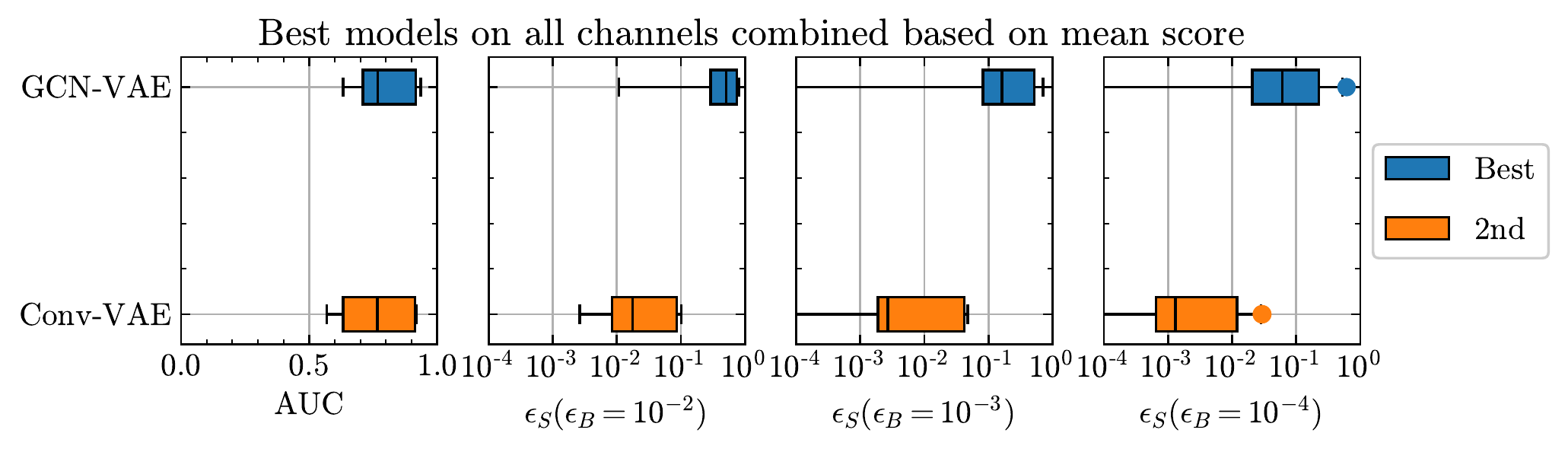}
    \caption{Comparison of the GCN-VAE and Conv-VAE performances, in terms of the benchmark figures of merit adopted in the paper.\label{fig:GCNvsConv}}
\end{figure}

Considering all comparison metrics along with the representative results shown in Fig.~\ref{fig:GCNvsConv}, graph architectures exhibit a definitive improvement in performance compared to the Conv-VAE. 
The improvement is seen not only in the AUC metric, but more significantly in the $\epsilon_\mathrm{S}$ at low $\epsilon_\mathrm{B}$. 
Because of this, the GCN-VAE is used as the reference architecture in the rest of this section and in section~\ref{sec:VAE_Flows}.\\

\subsection{Physics-motivated high-level features}
\label{sec:event_variables}

We also experiment with adding physics-motivated high-level features, as explicit inputs to the model, similar to what was done with object multiplicities in section~\ref{sec:data_representation}. 
Doing so, we intend to check if domain knowledge helps in improving anomaly detection capability. 
We pass event information such as the missing transverse momentum in the event ($\ptmiss$), the scalar sum of the jet $\pt$ ($\HT$) and $m_{\mathrm{Eff}}=\HT+\ptmiss$ to the model, by concatenating these with the output of the convolutional layers of the encoder. 
The concatenated output is then passed to the fully connected layers in the encoder to form the latent space. 
After the point sampled from the latent space passes through the fully connected layers of the decoder, the reconstructed $\ptmiss$, $\HT$ and $m_{\mathrm{Eff}}$ are extracted and the rest of the layer output is re-shaped and further passed to the subsequent layers of the decoder.

To include the reconstruction of these features in the loss, we add to Eq.~(\ref{eq:VAELoss}) a mean-squared error (MSE) term, computed from the reconstructed and input high-level features and weighted by a coefficient. 
This coefficient is treated as a hyperparameter that is scanned until the best performance is found.

\begin{figure}[t!]
    \centering
    \includegraphics[width=0.8\linewidth]{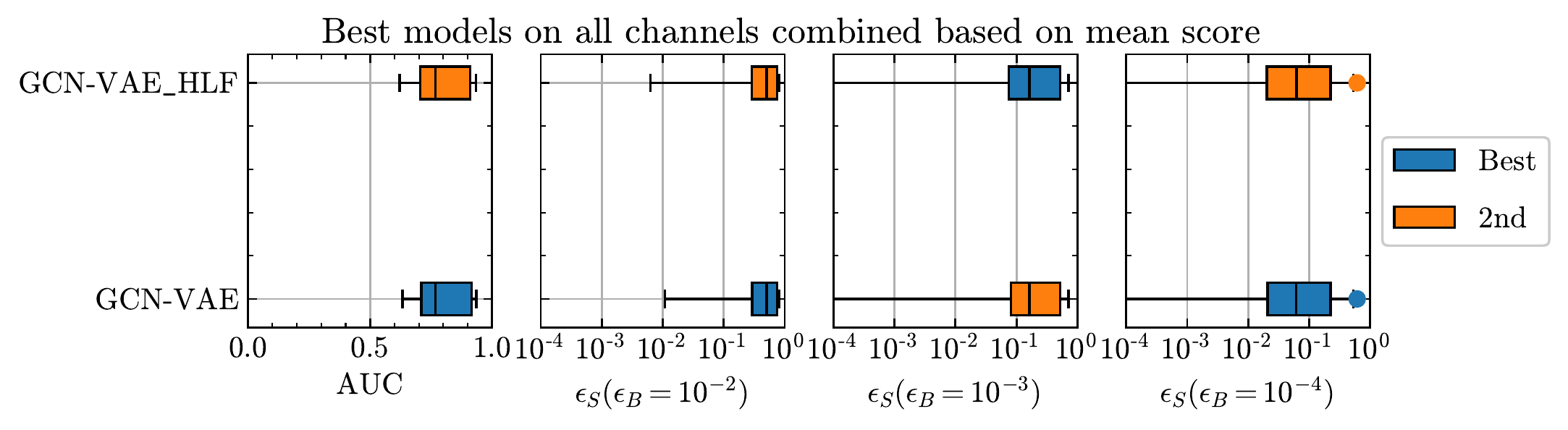}
    \caption{Comparison of the GCN-VAE performance with and without high-level features added as a separate input.
    \label{fig:HighLVLFeatures}}
\end{figure}

Figure~\ref{fig:HighLVLFeatures} shows that adding high-level features brings no definitive improvement in performance, thereby leading us to conclude that the baseline model with marginally lower number of trainable parameters is a good choice.\\

\subsection{Anomaly scores}
\label{sec:anomaly score}

While so far the Chamfer loss has been used as the anomaly score, this is not the only possibility. 
We consider two alternative metrics: the $D_\mathrm{KL}$ term in Eq.~(\ref{eq:VAELoss}) and~\citep{aarrestad2021dark}:
\begin{equation}
    R_z = \sum_i \left ( \frac{\mu_i}{\sigma_i} \right )^2
\end{equation}
where $\mu$ and $\sigma$ are the mean and RMS returned by the encoder and the index $i$ runs across the latent-space dimensions. 

The use of different anomaly scores requires a tuning of the $\beta$ hyperparameter. 
Since $\beta$ determines the relative importance of the $D_\mathrm{KL}$ and Chamfer loss terms in the loss, the use of one or the other as anomaly score is certainly related to the choice of the optimal $\beta$ value. 
Similarly, the use of $R_z$ (i.e., anomaly detection in the latent space) might not be optimal when using a $\beta$ value that was tuned to emphasize the reconstruction accuracy (i.e., the minimization of the Chamfer term in the loss). 
On the other hand, the study in \cite{aarrestad2021dark} shows that an excessive tuning of the hyperparameters affects generalization of performance negatively beyond the available dataset.

In order to address this point, we compare three  weights for the $\beta$ term. 
The first case ($\beta=1$) corresponds to training the VAE without the contribution of the reconstruction loss. 
In the second case ($\beta=0.5$) the two contributions are equally weighted. The final case ($\beta=10^{-6}$) corresponds to suppressing the $D_\mathrm{KL}$ term to a negligible level.

\begin{figure}[t!]
    \centering
    \includegraphics[width=\linewidth]{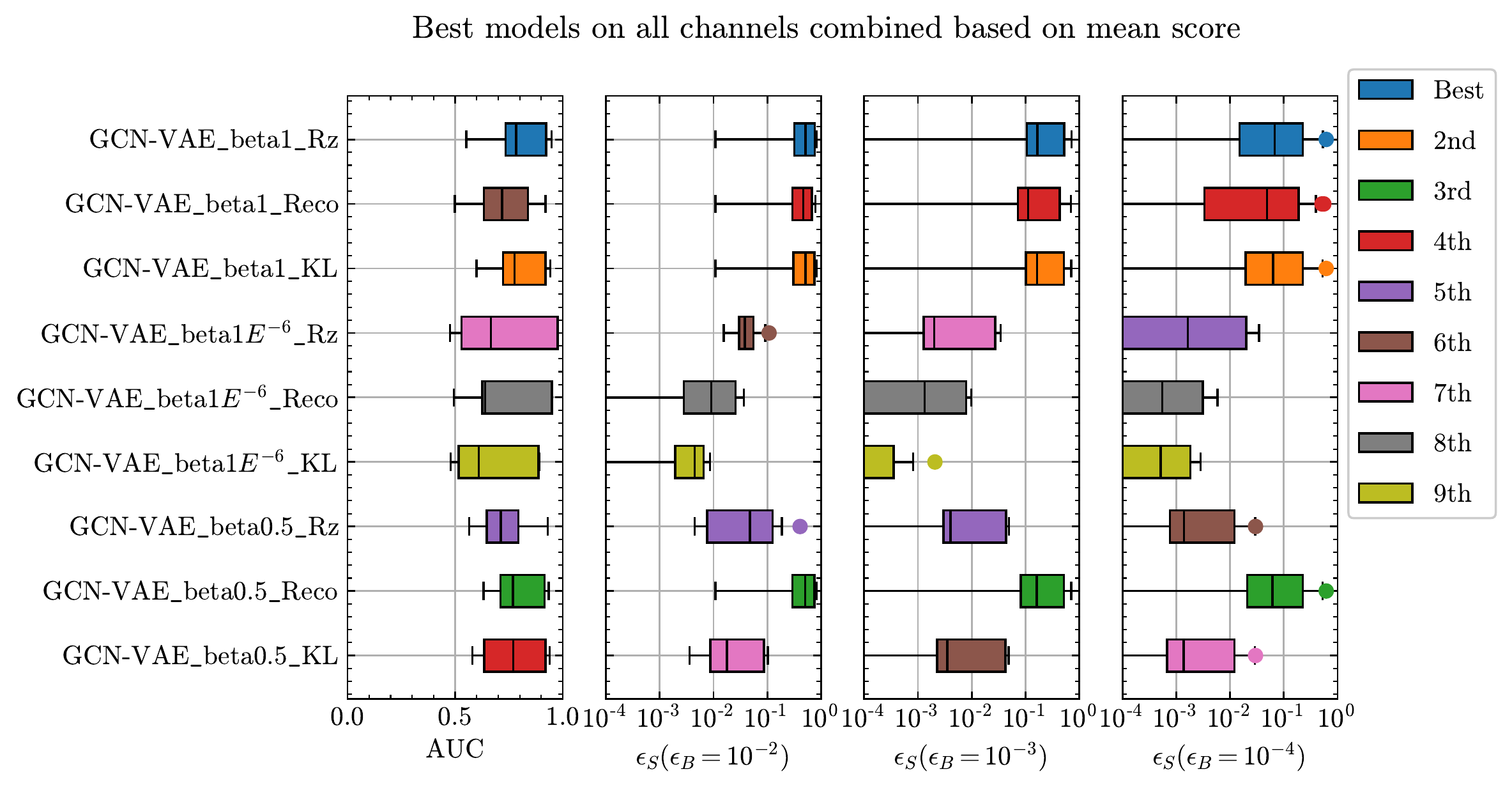}
    \caption{Comparison of anomaly detection performance from different anomaly score definitions, applied to the GCN-VAE.\label{fig:betaVAE}}
\end{figure}

Figure~\ref{fig:betaVAE} shows that all three anomaly scores underperform in the $\beta=10^{-6}$ case. 
The best performing models overall are the $\beta=1$ and $\beta=0.5$ cases. 
Comparing across the three different anomaly scores, we see that the $\beta=1$ model that uses $D_\mathrm{KL}$ and $R_z$ metrics, as well as the $\beta=0.5$ model that uses the reconstruction metric perform the best. 
All three cases also show very similar performance across all comparison metrics as well as methods, implying that either model-anomaly score combination is equally suitable. 
We also find that the  $\beta=1$ $D_\mathrm{KL}$ score and the $\beta=0.5$ reconstruction score show a similar correlation pattern on signal and background. 
As a result, we expect that only a limited improvement would be obtained by combining the two, which spares us the cost of introducing a new hyperparameter (the relative weight of the two terms) whose optimal value would be signal-specific, as in the case of \cite{Caron:2021wmq}.

\subsection{Baseline discrimination}

\begin{figure}[!t]
    \centering
    \includegraphics[width=0.48\linewidth]{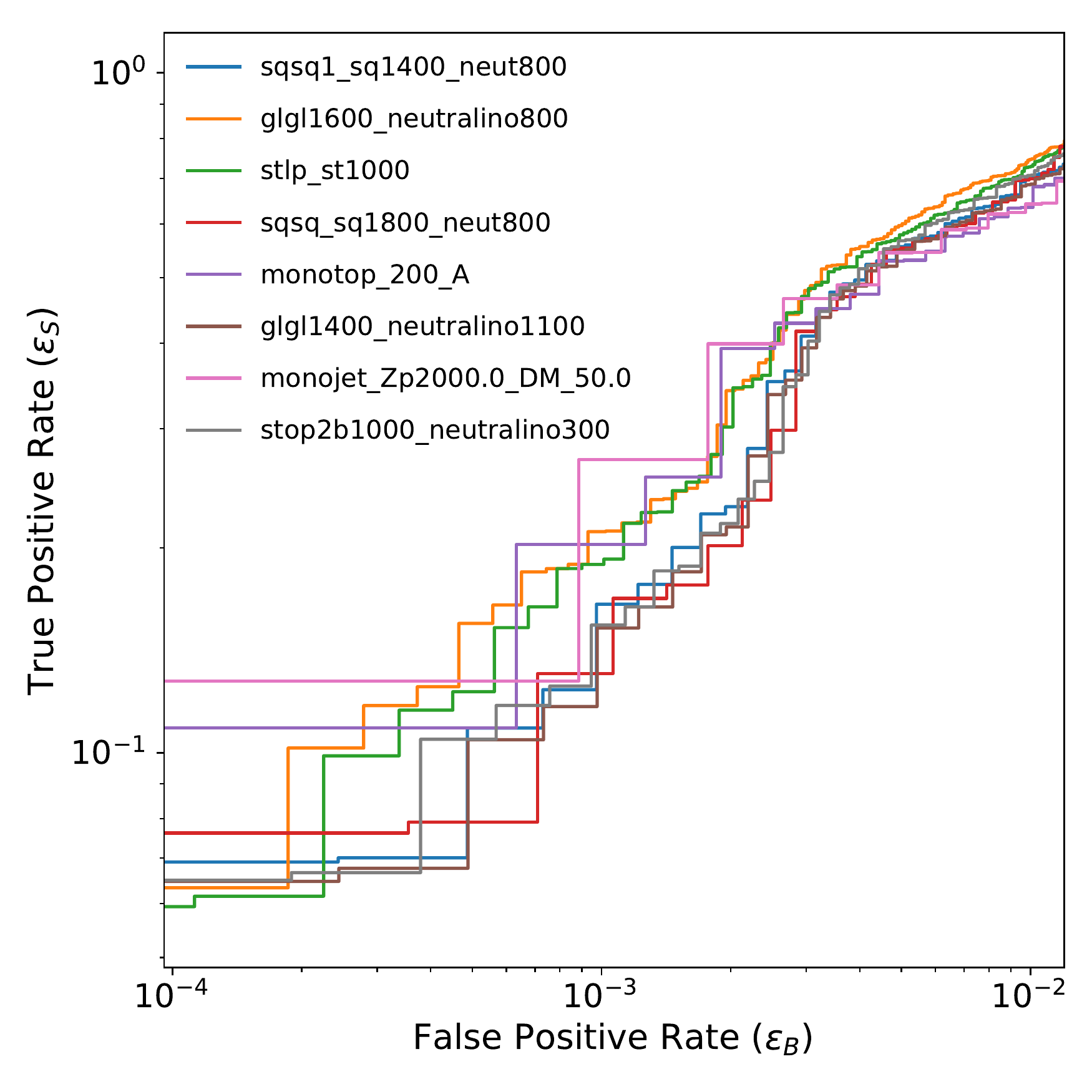}
    \includegraphics[width=0.48\linewidth]{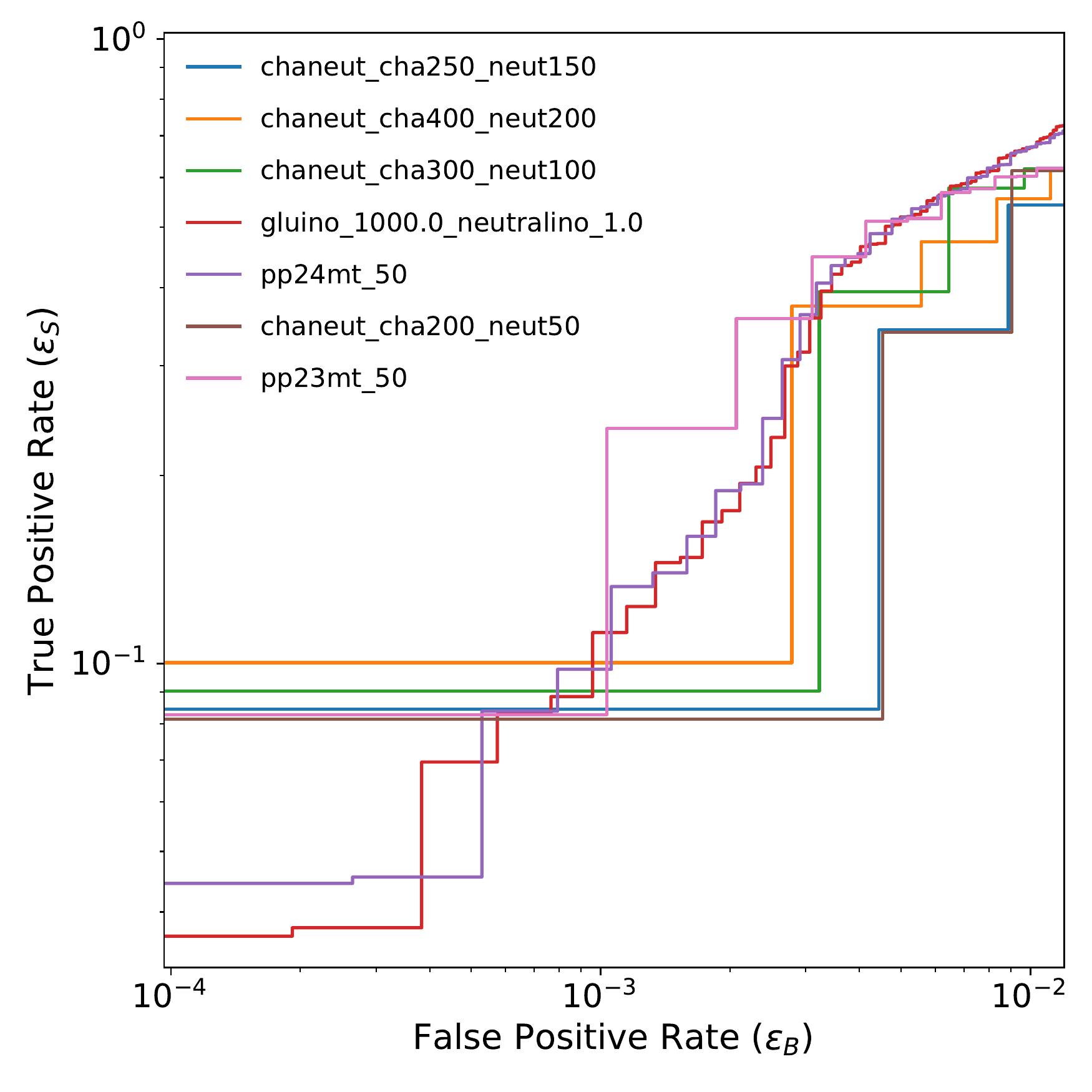} \\
    \includegraphics[width=0.48\linewidth]{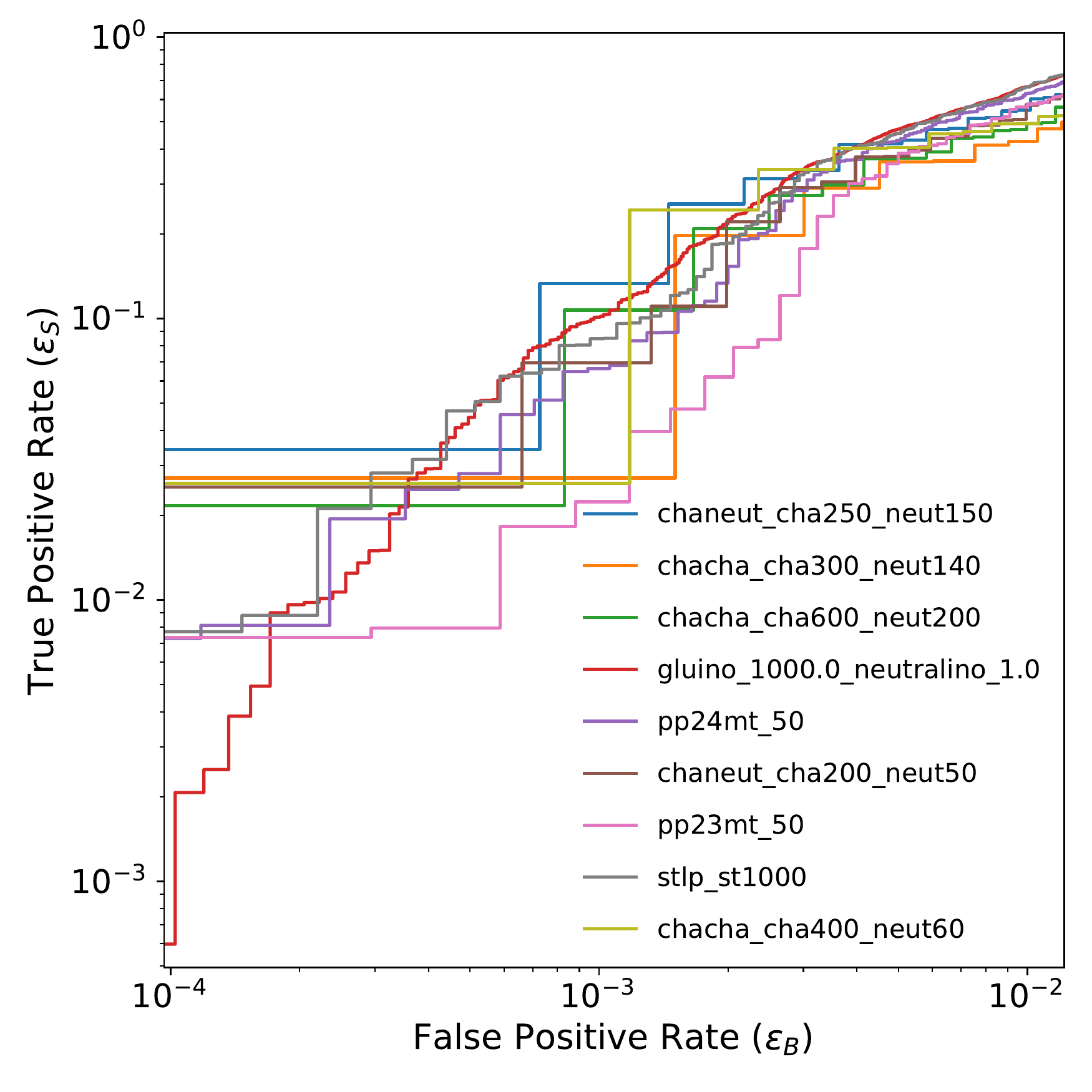}
    \includegraphics[width=0.48\linewidth]{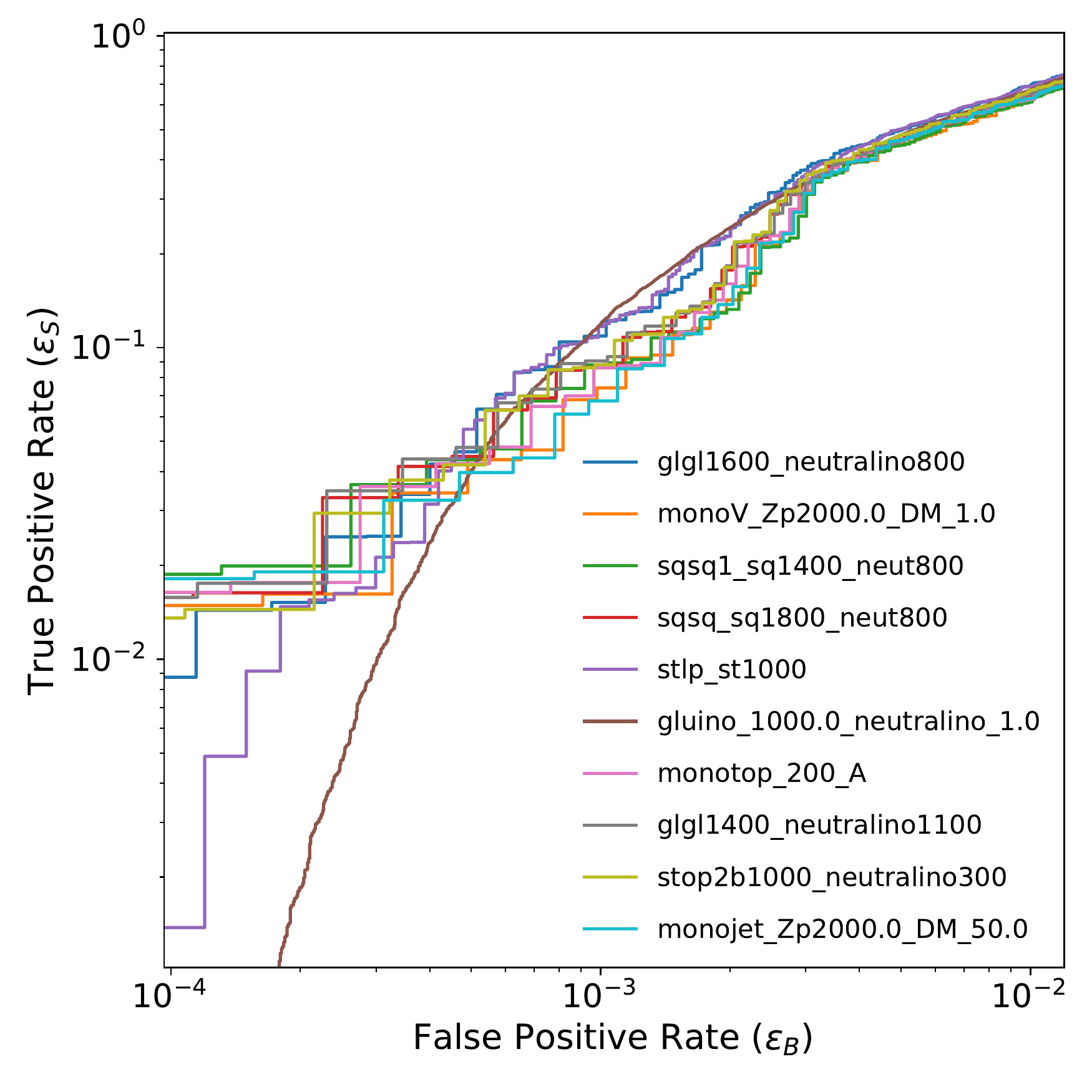}
    \caption{ROC curves for the baseline GCN-VAE model in channel 1 (top left), channel 2a (top right), channel 2b (bottom left), and channel 3 (bottom right), computed from the $\epsilon_\mathrm{S}$ and $\epsilon_\mathrm{B}$ values obtained on the background sample and the benchmark signal samples. 
    Most of the ROC curves are not smooth, due to the small dataset size for some of the channels.\label{fig:ADscore_ROC}}
\end{figure}

As a result of the tests presented so far, the baseline VAE model is established as a GCN-VAE taking as input the whole set of reconstructed physics object but no domain-specific high level features. 
The Chamfer loss function is used as the anomaly score. 
The GCN-VAE is trained and tested only with data available within a given channel and the dataset sizes per channel are described in Table~\ref{tab:dataset_size}. 
Figure~\ref{fig:ADscore_ROC} shows the ROC curves for the baseline VAE model on benchmark signals in the four channels. 
It is evident that we suffer from a shortage of events for some signal models at very low $\epsilon_\mathrm{B}$.
We still show ROC curves down to $\epsilon_\mathrm{B}=10^{-4}$ to allow one to compare our results to those in \cite{aarrestad2021dark}, where this range was chosen.
We see an overall improvement in $\epsilon_\mathrm{S}$ at very low $\epsilon_\mathrm{B}$ for the GCN-VAE compared to our Conv-VAE submission in~\cite{aarrestad2021dark}.\\

\section{Normalizing flows}
\label{sec:VAE_Flows}
With the GCN-VAE serving as the baseline, we investigate how the use of NFs~\citep{tabak2010density,tabak2013family} impacts the anomaly-detection performance. 
Normalizing flow layers are inserted between the Gaussian sampling and the decoder. 
They provide additional complexity to learn better posterior distributions~\citep{rezende2015variational} by morphing the multivariate prior of the latent space to a more suitable, learned function. 

In other words, we use the NF layers to handle the fact that a VAE converging to a good output-to-input matching does not necessarily correspond to a configuration with a Gaussian prior in the latent space, $p(z)=\prod G(z)$. 
To reach this configuration (e.g., when training a VAE as a generative model), one typically uses a $\beta$-VAE with an increased weighting of the $D_\mathrm{KL}$ regularizer. 
This typically results in a degradation of the output-to-input matching. 
With NFs, we learn a generic prior $p(z)$ as $f(G(z))$, where $f$ is the transformation function learned by the NF layers. 
This is different from the way NFs are traditionally used in VAE training, i.e., to improve the convergence of $f(z)$ to $G(z)$ with a stronger evidence lower bound (ELBO) condition. 
Because of this, we do not modify the $D_\mathrm{KL}$ term in the loss, as done in~\cite{rezende2015variational}. 
The results obtained following this more traditional training procedure are described in the supplementary material. 
Doing so, we observe worse $\epsilon_\mathrm{S}$ for the same $\epsilon_\mathrm{B}$. 
This is expected because the ELBO improvement with NFs was introduced in~\cite{tomczak2017improving} as a way to improve the VAE generative properties, and it does not imply a better anomaly detection capability.

A NF can be generalized as any invertible, diffeomorphic transformation that can be applied to a given distribution to produce tractable distributions~\citep{papamakarios2021normalizing, kobyzev2020normalizing}. 
In order to be compatible with variational inference, it is desirable for the transformations to have an efficient mechanism for computing the determinant of the Jacobian, while being invertible~\citep{rezende2015variational}. 
The NFs are trained sequentially, together with the baseline VAE model.

We utilize four major families of flow models:
\begin{itemize}

    \item \textbf{Planar flows} are invertible transformations whose Jacobian determinant can be computed rather efficiently, making them suitable candidates for variational inference~\citep{rezende2015variational}. 
    PF transformations are defined as:
    \begin{equation}
      \textbf{z}' = \textbf{z} + \text{u}h(\text{w}^T\textbf{z} + b)~~,
    \end{equation}
    where  $\text{u}, \text{w} \in \mathbb{R}^D$, $b \in \mathbb{R}$ and $h$ is a suitable smooth activation function. 

    \item \textbf{Sylvester normalizing flows} (SNFs)~\citep{berg2018sylvester} build on the planar flow formulation and extend it to be analogous to a multilayer perceptron with one hidden layer of $M$ units and a residual connection as:
    \begin{equation}
        \textbf{z}' = \textbf{z} + \textbf{A}h(\textbf{B}\textbf{z} + b)~~,
    \end{equation}
    where  $\text{A} \in \mathbb{R}^{D\times M}, \text{B} \in \mathbb{R}^{M\times D}$, $b \in \mathbb{R}^M$ and $M\leq D$. 
    Computing the Jacobian determinant for such a formulation is made more efficient by utilizing the Sylvester determinant identity~\citep{berg2018sylvester}. 
    Depending on the way $A$ and $B$ are parametrized, we get different types of SNFs. 
    In this paper we use orthogonal, Householder, and triangular SNFs, as described in \cite{berg2018sylvester}.

    \item \textbf{Inverse autoregressive flows} (IAFs)~\citep{kingma2016improved} are computation-efficient normalizing flows based on autoregressive models. 
    Autoregressive transformations are invertible, making them suitable candidates for our case. 
    However, computing the transformation requires multiple sequential steps~\citep{berg2018sylvester}. 
    The inverse transformation however, leads to certain simplifications as described in \cite{berg2018sylvester}, allowing more efficient parallel computing, thereby making it a more desirable transformation for our case. 
    We use the IAFs  formulated as:
    \begin{equation}
        z_i^t = \mu_i^t(z_{1:i-1}^{t-1}) + \sigma_i^t(z_{1:i-1}^{t-1}) \cdot z_i^{t-1} \quad , \quad i = 1, 2, ..., D~.
    \end{equation}
    Such a formulation allows one to stack multiple transformations to achieve more flexibility in producing target distributions.

    \item {\bf Convolutional normalizing flows} (ConvolutionalFlows)~\citep{zheng2017convolutional} are an extension of single-hidden-unit planar flows~\citep{kingma2016improved} to the case of multiple hidden units, further enhanced by replacing the fully connected network operation with a one-dimensional (1D) convolution, to achieve bijectivity.
    They are defined by the following transformation:
    \begin{equation}
        \textbf{z}' = \textbf{z} + \textbf{u}\odot h(\text{conv}(\textbf{z},\textbf{w}))~~,
    \end{equation}
    where $w \in R^k$ is the parameter of the 1D convolution filter with $k$-sized kernel, $h$ is a monotonic nonlinear activation function and $\odot$ denotes pointwise multiplication.
    
    \item {\bf Autoregressive neural spline flows} (NSFARs)~\citep{durkan2019neural} are similar to IAFs, where affine transforms are replaced by monotonic rational-quadratic spline transforms as described in~\cite{durkan2019neural}.  They resemble a traditional feed-forward neural network architecture, alternating between linear transformations and elementwise non-linearities, while retaining an exact, analytic inverse.   
\end{itemize}
 
The hyperparameters for each normalizing flow architecture are chosen arbitrarily to avoid overtuning on the available dataset as learned from \cite{aarrestad2021dark}. 
The planar flow model consists of a stack of six flows, each made of three dense layers with 90 neurons each. 
SNFs are defined by stacking six flows with 8 orthogonal, householder and triangular transformations for each of the respective types of SNF. 
IAFs are constructed with four masked autoencoder for distribution
estimation (MADE)~\citep{germain2015made} layers as described in~\cite{kingma2016improved}, each containing 330 neurons. 
ConvolutionalFlows include four flow layers with kernel size $k=7$ and applying kernel dilation as described in~\cite{zheng2017convolutional}. 
NSFARs are defined by stacking four flow layers each with $K=64$ bins and eight hidden features.

\begin{figure}[t!]
    \centering
    \includegraphics[width=0.8\linewidth]{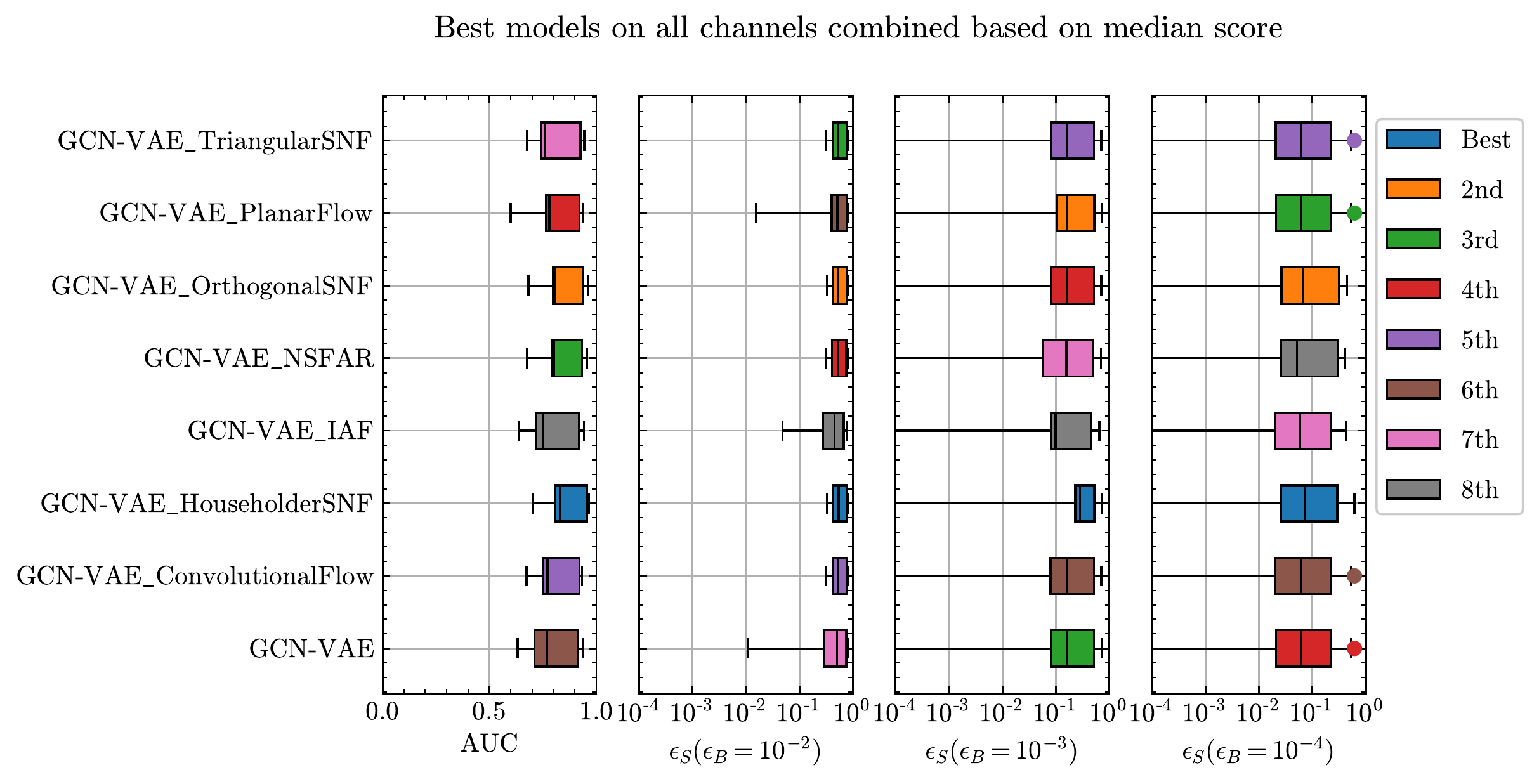}
    \caption{Comparison of anomaly detection performance for GCN-VAE models with different normalizing flow architectures in the latent space
    \label{fig:GCN_Flows}}
\end{figure}

Figure~\ref{fig:GCN_Flows} shows the results of all GCN-VAE models combined with all the different types of flows as described in section~\ref{sec:VAE_Flows}.  
Based on results from all data channels combined through all six strategies mentioned in section~\ref{sec:training_setup_metrics}, and considering variance across trainings from different random seeds (see supplementary material), it is evident that using normalizing flows improves not only the AUC metric but also the signal efficiencies at low background efficiencies.  
We find that the Householder variant of SNFs produces the best improvement with respect to the baseline GCN-VAE model. 
The exercise was also repeated with a Conv-VAE model and similar trends were observed. 
There, the normalizing flows showed a larger improvement from the baseline Conv-VAE than for the GCN-VAE model but the overall results are less accurate than that of GCN-VAE with normalizing flows.

\begin{figure}[ht!]
    \centering
    \includegraphics[width=0.48\linewidth]{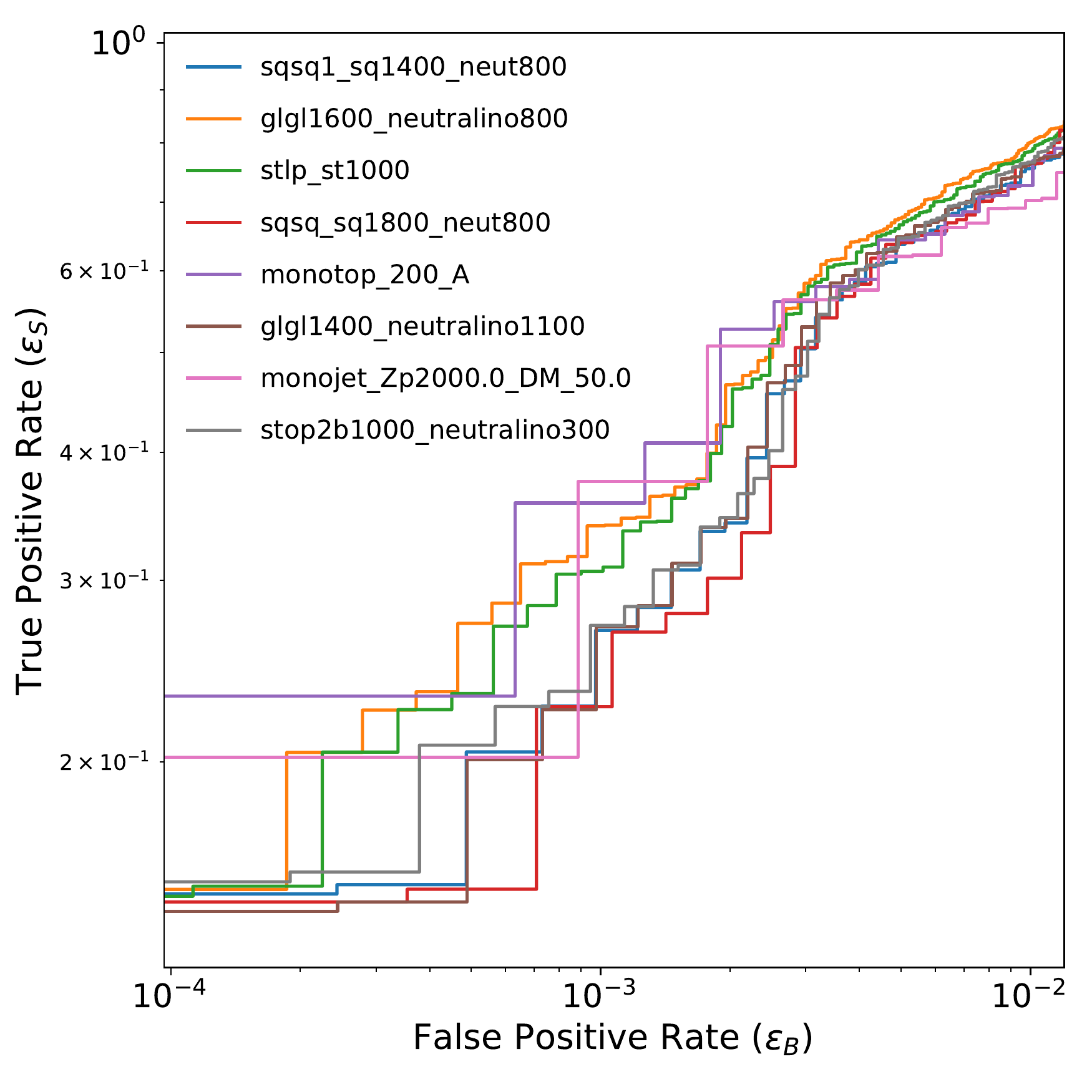}
    \includegraphics[width=0.48\linewidth]{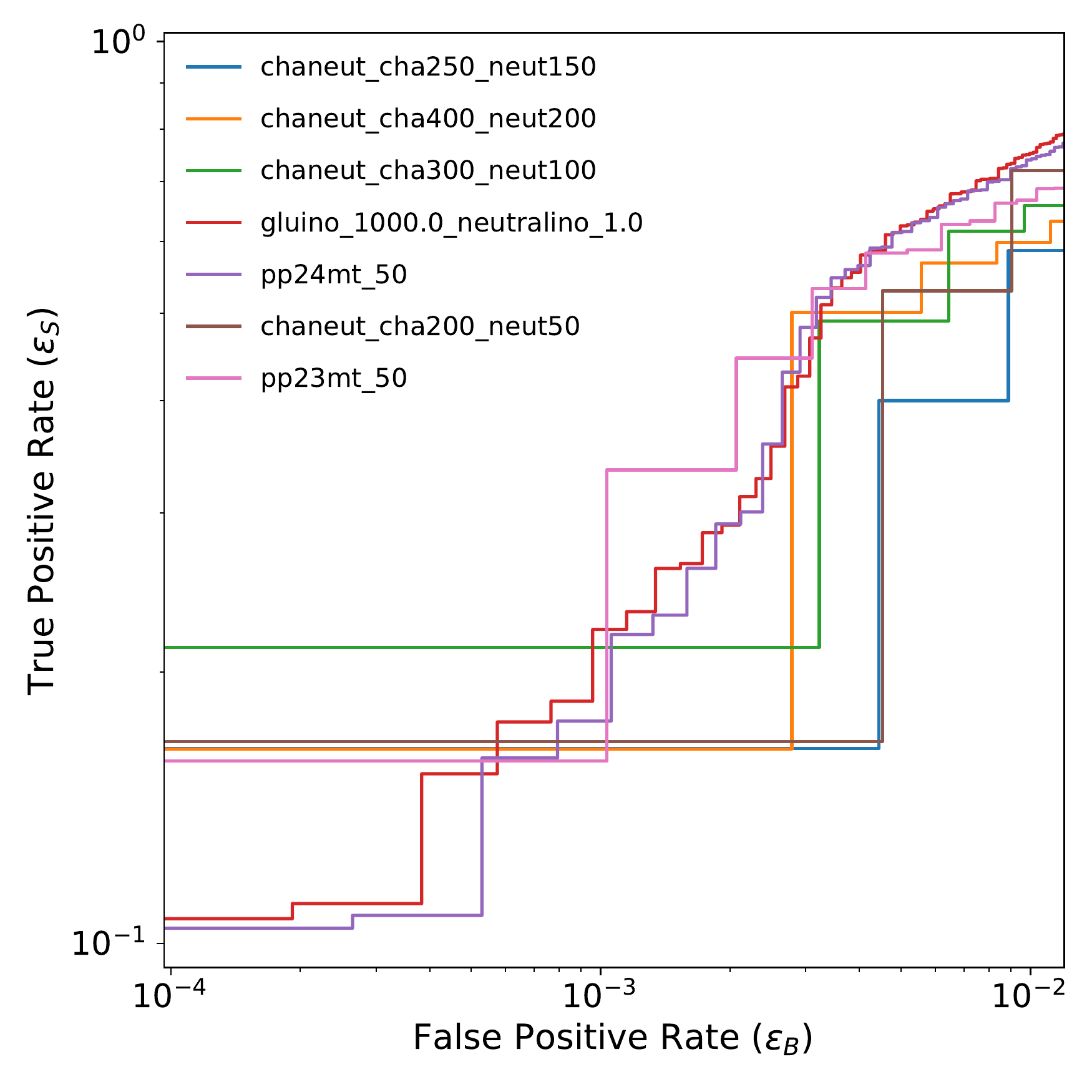}\\
    \includegraphics[width=0.48\linewidth]{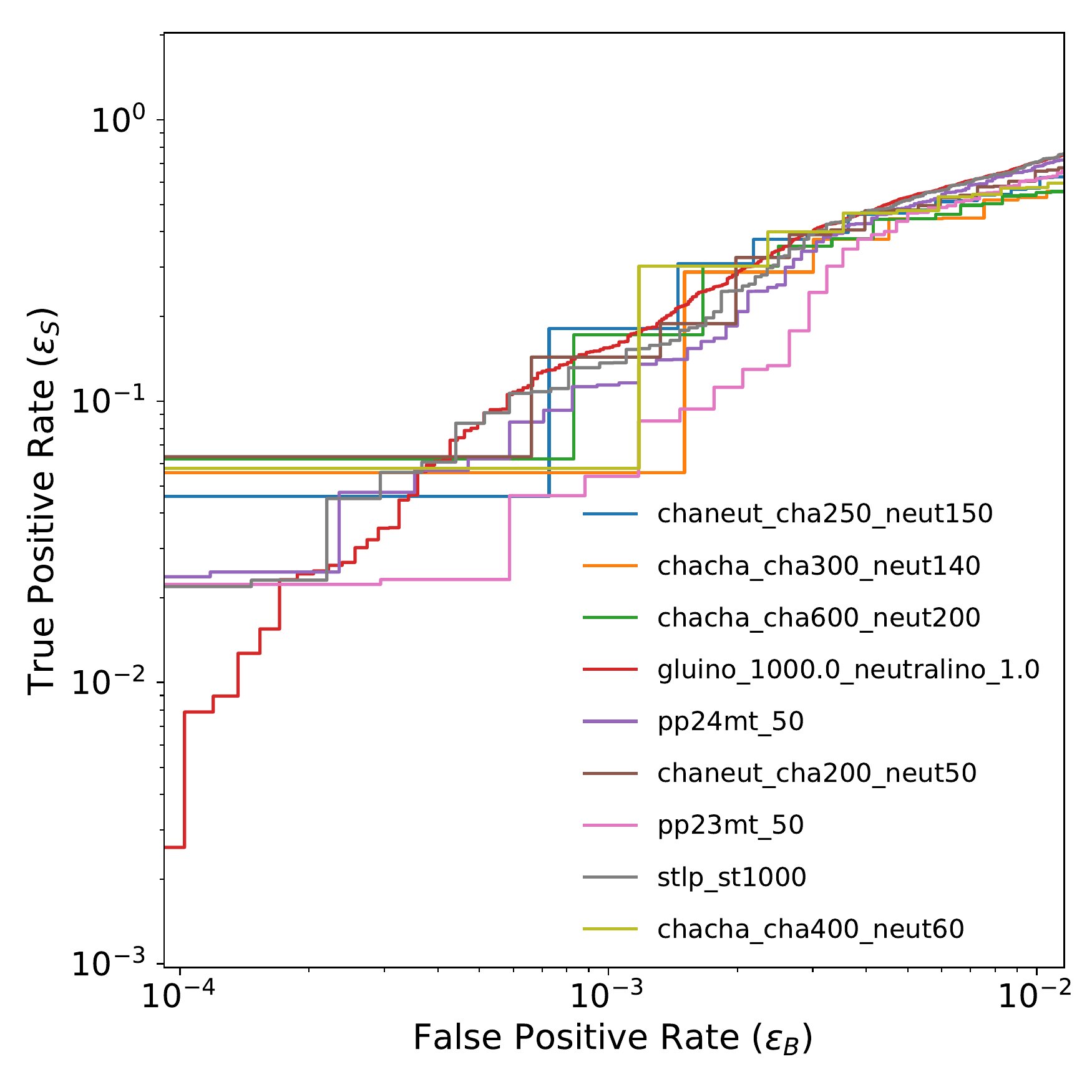}
    \includegraphics[width=0.48\linewidth]{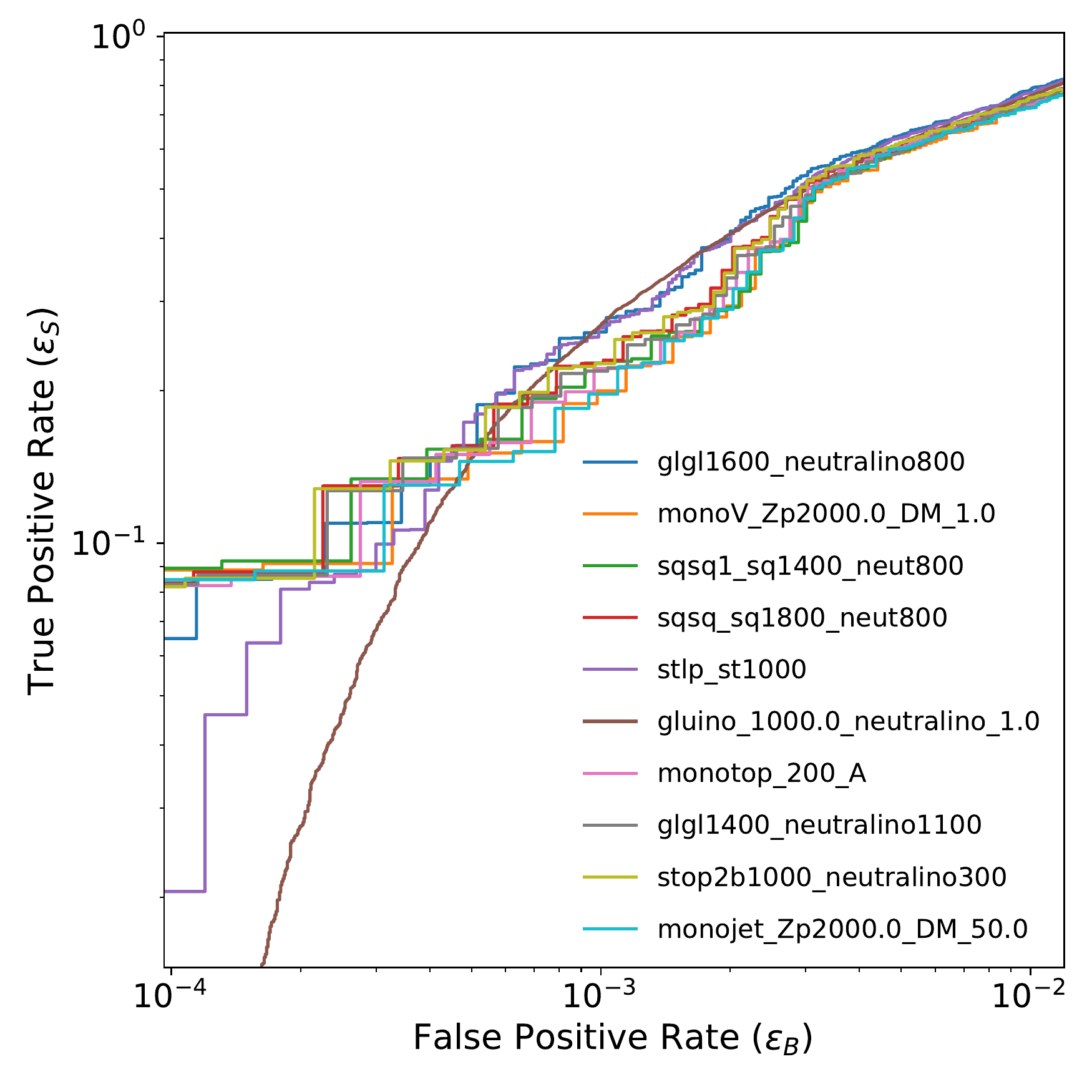}
    \caption{ROC curves of GCN-VAE\_HouseholderSNF for all signals in each of channel 1 (top left), channel 2a (top right), channel 2b (bottom left), and channel 3 (bottom right).
    \label{fig:bestModelROCs}}
\end{figure}

Figure~\ref{fig:bestModelROCs} shows the ROC curves for the best presented model, GCN-VAE\_HouseholderSNF across all available signal samples in all data channels. 
For some of the samples, the small dataset size translates in a discontinuous curve and larger uncertainties. 

\section{Conclusions}

We constructed a graph-based anomaly detection model to identify new physics events in the DarkMachines challenge dataset. 
Inspired by the outcome of this challenge, specific model design choices (data representation, use of physics-motivated high-level features, and anomaly score definition) were further optimized in order to maximize anomaly detection performance. 
As the case for many other deep learning applications to particle-physics data, we observed that the graph architecture better captures the point-cloud nature of this data, resulting in an enhanced performance.

In this baseline, we investigate the impact of using a stack of normalizing flows in the latent space of the variational autoencoder (VAE), between the Gaussian sampling and the decoding, in order to improve the accuracy of the prior learning process, by morphing the Gaussian prior to a more suitable function, learned during the training. 

Testing the trained model on a set of benchmark signal samples, we observe an overall improvement when normalizing flows are used, with the Householder variant of the Sylvester normalizing flow model giving the best results. 
With that, we reach a median anomaly identification probability of 72\% (34\%) for an $\epsilon_\mathrm{B}$ of 1\% (0.1\%) across all signal samples over all available channels. The median anomaly identification probability increases to 95\% (96\%) for an $\epsilon_\mathrm{B}$ of 30\% (60\%). 

This work presents an improvement over our Conv-VAE model, submitted to the DarkMachines challenge~\citep{aarrestad2021dark}.



\section*{Funding}
P.~J., T.~A., M.~P., and K.~A.~W. are supported by the European Research Council (ERC) under the European Union's Horizon 2020 research and innovation program (Grant Agreement No. 772369).
J.~D. is supported by the U.S. Department of Energy (DOE), Office of Science, Office of High Energy Physics Early Career Research program under Award No. DE-SC0021187.
S.~T. was supported by the University of California San Diego Triton Research and Experiential Learning Scholars (TRELS) program.
J.~N. is supported by Fermi Research Alliance, LLC under Contract No. DE-AC02-07CH11359 with the U.S. Department of Energy, Office of Science, Office of High Energy Physics.


\bibliographystyle{frontiersSCNS}
\bibliography{references.bib}

\noindent\textbf{Conflict of Interest}: The authors declare that the research was conducted in the absence of any commercial or financial relationships that could be construed as a potential conflict of interest.

\ifarxiv 
\clearpage
\setcounter{section}{0}
\setcounter{figure}{0}
\setcounter{table}{0}
\renewcommand{\thefigure}{S\arabic{figure}}
\renewcommand{\thetable}{S\arabic{table}}
{\huge\helveticaitalic{\textbf{Supplementary Material}}}
\section{Model variance}
\label{appendix:a}
In general, deep learning models are bound to have a variance in the results arising from different trainings with different random seeds for the stochastic gradient descent optimization method. 
To ensure that does not affect our inferences based on the final anomaly identification results, we performed five separate trainings with different random seeds. 
Table~\ref{tab:results_varianceTest} summarizes the median anomaly identification performance across all channels with the added variance across these five separate trainings, for the Conv-VAE model, the baseline GCN-VAE, and the GCN-VAE with the addition of the various normalizing flow models. 
Compared to the observed uncertainties, the improvements discussed in the paper are statistically significant.

\begin{table}[ht!]
    \centering
    \caption{Anomaly detection performance across all channels, combined based on median scores along with the variance over multiple trainings.}
    \label{tab:results_varianceTest}
    \begin{tabular}{c|c|c|c|c}
        Model & AUC & $\epsilon_\mathrm{S}(\epsilon_\mathrm{B}=10^{-2})$ & $\epsilon_\mathrm{S}(\epsilon_\mathrm{B}=10^{-3})$ & $\epsilon_\mathrm{S}(\epsilon_\mathrm{B}=10^{-4})$\\
        \hline
        
        Conv-VAE & $75.6\%\pm 0.5\%$ & $1.7\%\pm 0.6\%$ & $0.26\%\pm 0.04\%$ & $0.13\%\pm 0.07\%$ \\
        GCN-VAE & $76.8\%\pm 0.2\%$ & $55.7\%\pm 1.1\%$ & $16.2\%\pm 0.6\%$ & $7.1\%\pm 0.1\%$ \\
        \bf GCN-VAE\_HouseholderSNF & $\bm{86.1\%\pm 0.4\%}$ & $\bm{69.6\%\pm 1.9\%}$ & $\bm{34.2\%\pm 0.9\%}$ & $\bm{8.2\%\pm 0.5\%}$ \\
        GCN-VAE\_OrthogonalSNF & $82.4\%\pm 0.6\%$ & $65.0\%\pm 1.4\%$ & $16.1\%\pm 1.4\%$ & $7.6\%\pm 0.1\%$ \\
        GCN-VAE\_NSFAR & $82.1\%\pm 0.4\%$ & $64.1\%\pm 1.8\%$ & $15.6\%\pm 1.0\%$ & $5.1\%\pm 0.3\%$ \\
        GCN-VAE\_PlanarFlow & $80.0\%\pm 0.7\%$ & $61.1\%\pm 1.4\%$ & $16.2\%\pm 1.2\%$ & $7.1\%\pm 0.1\%$ \\
        GCN-VAE\_ConvolutionalFlow & $79.2\%\pm 0.2\%$ & $62.7\%\pm 1.4\%$ & $16.0\%\pm 1.4\%$ & $6.0\%\pm 0.2\%$ \\
        GCN-VAE\_TriangularSNF & $75.9\%\pm 0.5\%$ & $64.7\%\pm 1.5\%$ & $16.1\%\pm 1.3\%$ & $6.1\%\pm 0.6\%$ \\
        GCN-VAE\_IAF & $75.1\%\pm 0.8\%$ & $59.2\%\pm 2.7\%$ & $9.8\%\pm 0.8\%$ & $5.8\%\pm 0.7\%$
        
\end{tabular}
\end{table}

\section{Choice of loss function}
\label{appendix:b}

The optimization function used for normalizing flow models in variational inference is commonly formulated as the free energy bound~\citep{rezende2015variational}:
\begin{equation}
    \label{eq:FreeEnergyBound}
    \mathcal{F}(x) = \mathbb{E}_{q_0(z_0)}[\ln q_0(z_0)]-\mathbb{E}_{q_0(z_0)}[\log q(x,z_K)]-\mathbb{E}_{q_0(z_0)}\Big[\sum_{k=1}^{K}\log \big|\mathrm{det}(J) \big|\Big]
\end{equation}
where $\log|\mathrm{det}(J)|$ stand for the log-det-Jacobian term~\citep{rezende2015variational} for the corresponding flow model. 
In our setup, this term would be added to the task-specific reconstruction loss. 
For our GCN-VAE\_HouseholderSNF model, we compare using this loss function with the added Chamfer loss described in Eq.~(\ref{eq:ChamferLoss}), against using the same loss function for our baseline VAE model as described in Eq.~(\ref{eq:VAELoss}). 
All other model parameters are kept identical between the two trainings. 
Table~\ref{tab:lossStudyResults} shows the results of this comparison with ``Modified ELBO" signifying Eq.~\ref{eq:FreeEnergyBound} with the added Chamfer term, and ``Our Loss" signifying Eq.~(\ref{eq:VAELoss}). 
It is evident that not modifying the loss for the VAE after adding the flow layers results in significantly better anomaly identification performance and as a result we utilize this strategy to train all other flow models presented in this study.

\begin{table}[ht!]
    \centering
    \caption{Anomaly detection performance across all channels, combined based on median scores.}
    \label{tab:lossStudyResults}
    \begin{tabular}{c|c|c|c|c}
        Model & AUC & $\epsilon_\mathrm{S}(\epsilon_\mathrm{B}=10^{-2})$ & $\epsilon_\mathrm{S}(\epsilon_\mathrm{B}=10^{-3})$ & $\epsilon_\mathrm{S}(\epsilon_\mathrm{B}=10^{-4})$\\
        \hline
        
        Modified ELBO & $78.3\%$ & $57.1\%$ & $16.1\%$ & $6.6\%$ \\
        \bf Our Loss & $\bm{86.4}\%$ & $\bm{71.6}\%$ & $\bm{34.0}\%$ & $\bm{8.2}\%$ \\
        
\end{tabular}
\end{table}

\section{Impact of normalizing flows on the latent space prior}
\label{appendix:c}
To understand exactly how non-Gaussian the latent distributions become after passing through the normalizing flow layers, we attempt to visualize our $15$ dimensional latent space via two methods. 
First, we create $2$D histograms from multiple randomly chosen pairs of dimensions. Fig.~\ref{fig:latentSpaceHist} shows one such distribution between dimensions $5$ and $9$. We also make an approximate visualisation by first performing a principal component analysis (PCA) to express the latent space in $2$ dimensions, and then plotting the resulting $2$D histogram as shown in Fig.~\ref{fig:latentSpacePCA}. 

\begin{figure}[t!]
    \centering
    \includegraphics[width=0.38\linewidth]{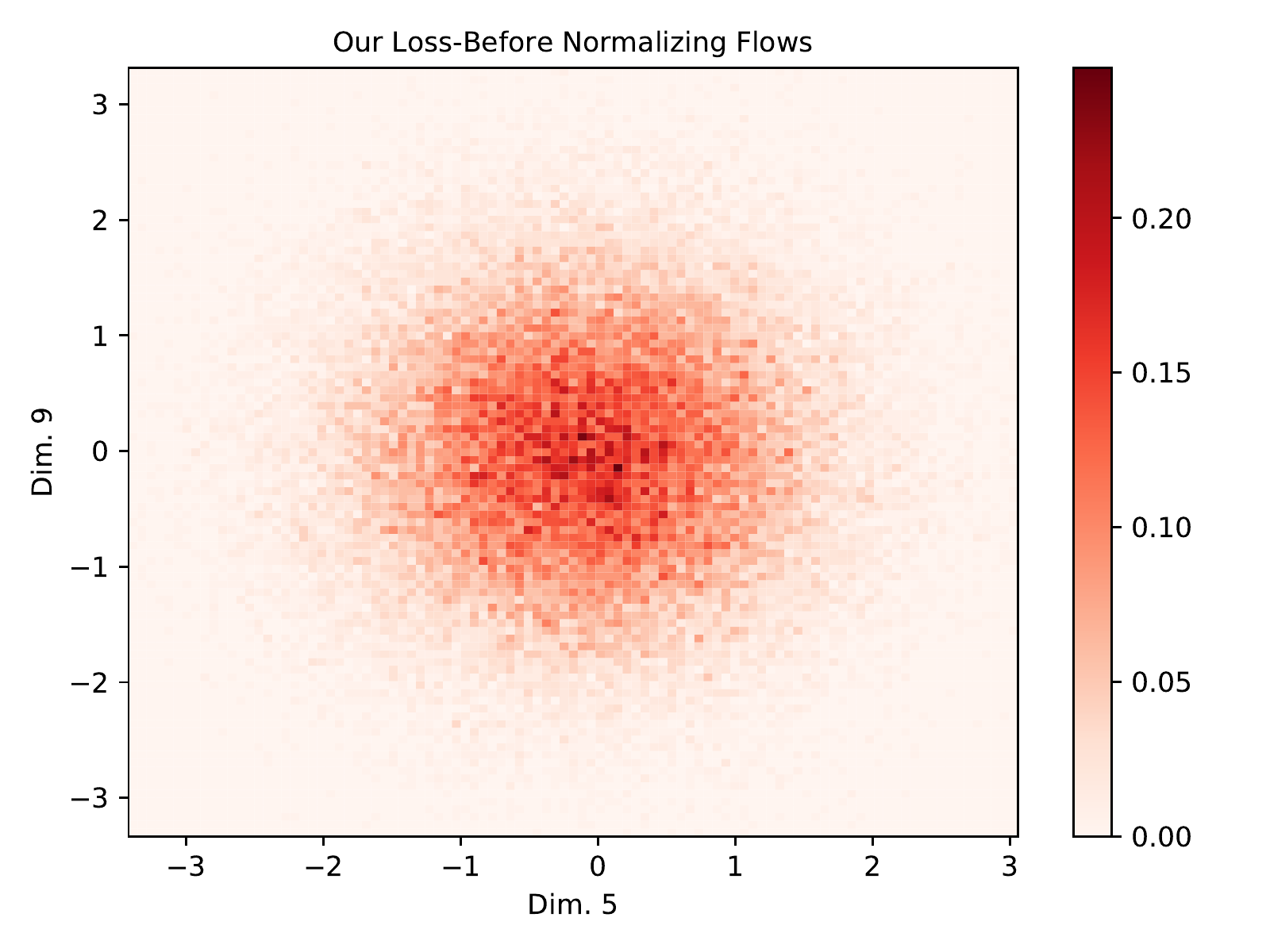}
    \includegraphics[width=0.38\linewidth]{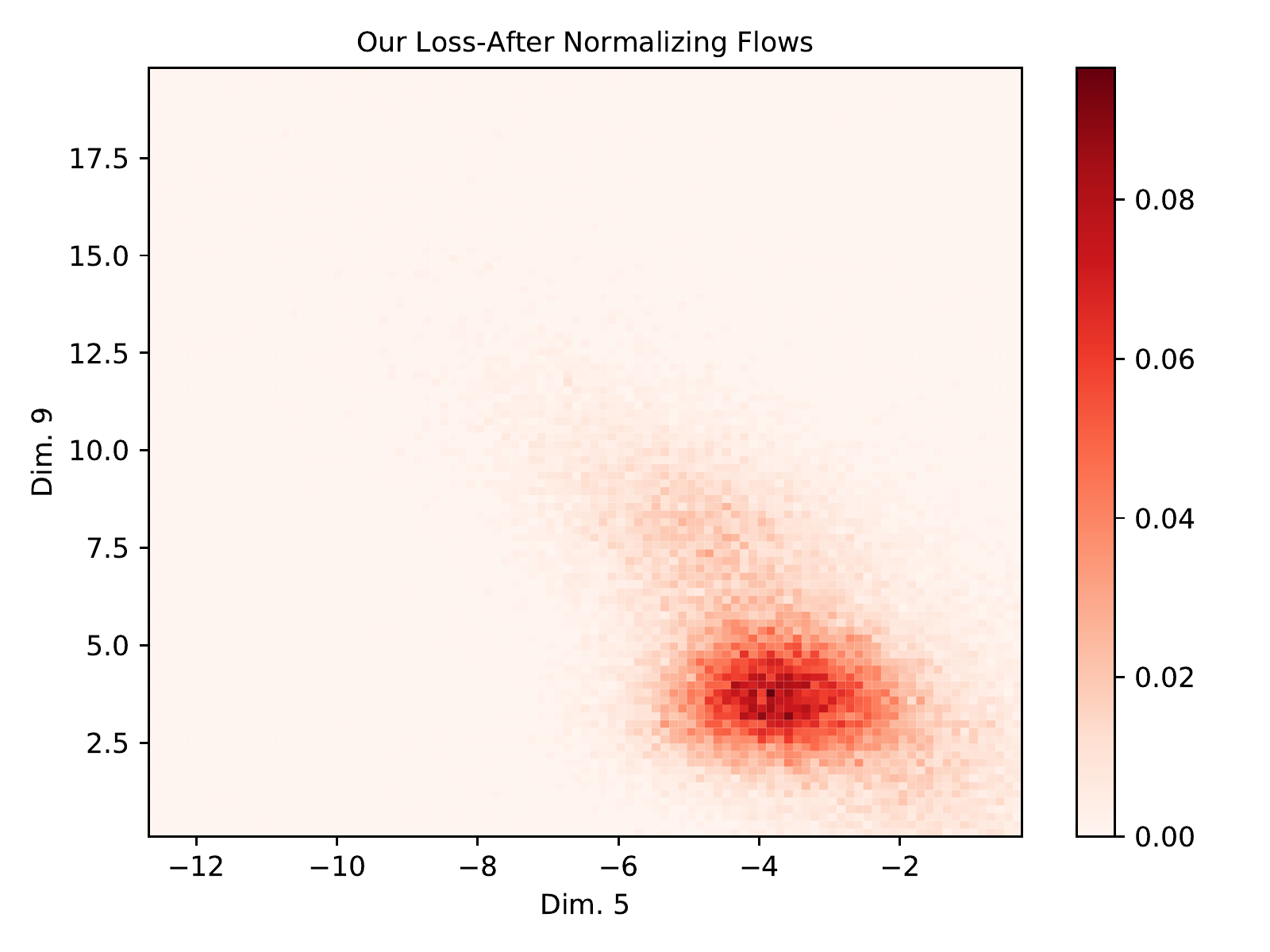}\\
    \includegraphics[width=0.38\linewidth]{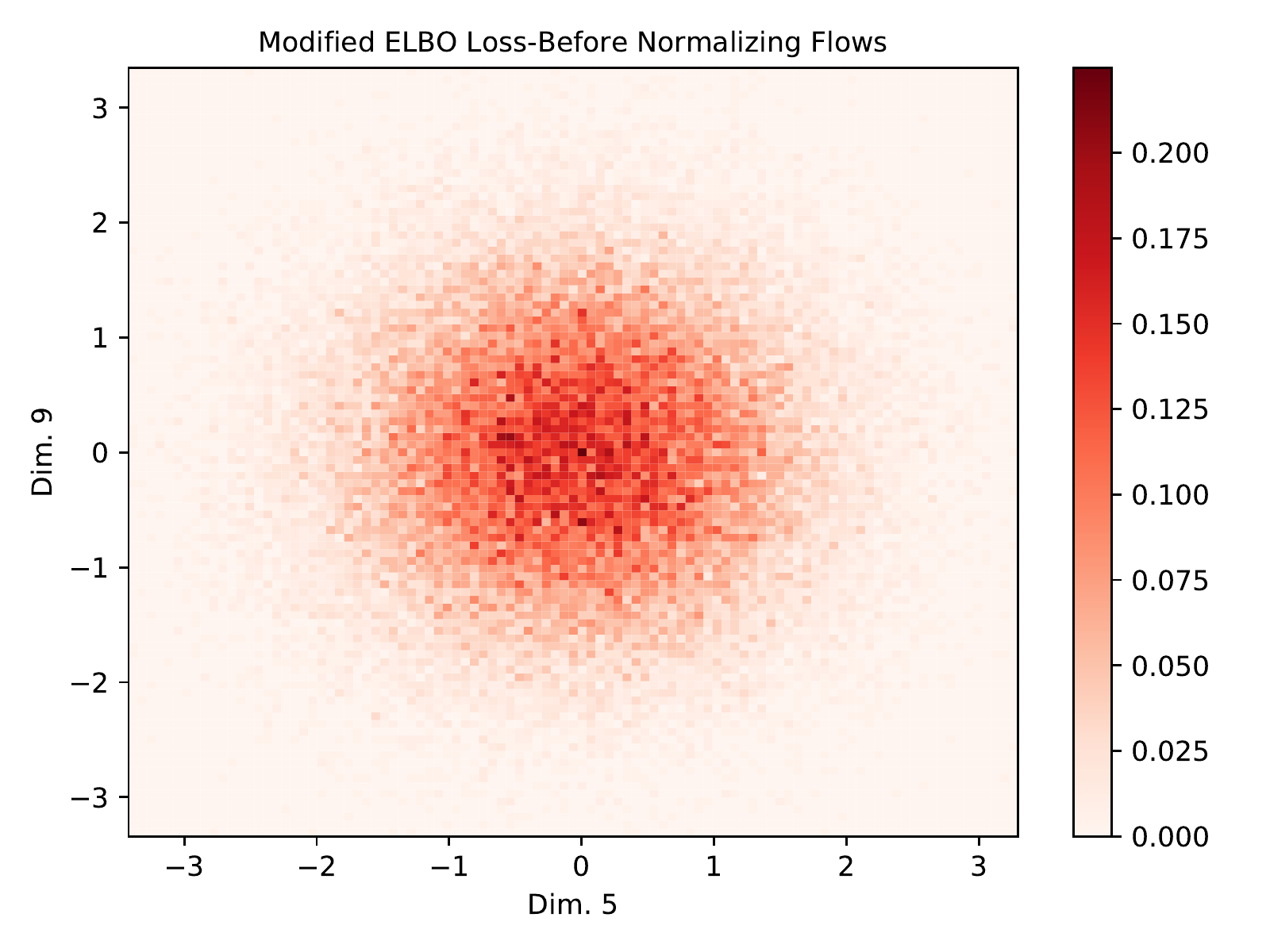}
    \includegraphics[width=0.38\linewidth]{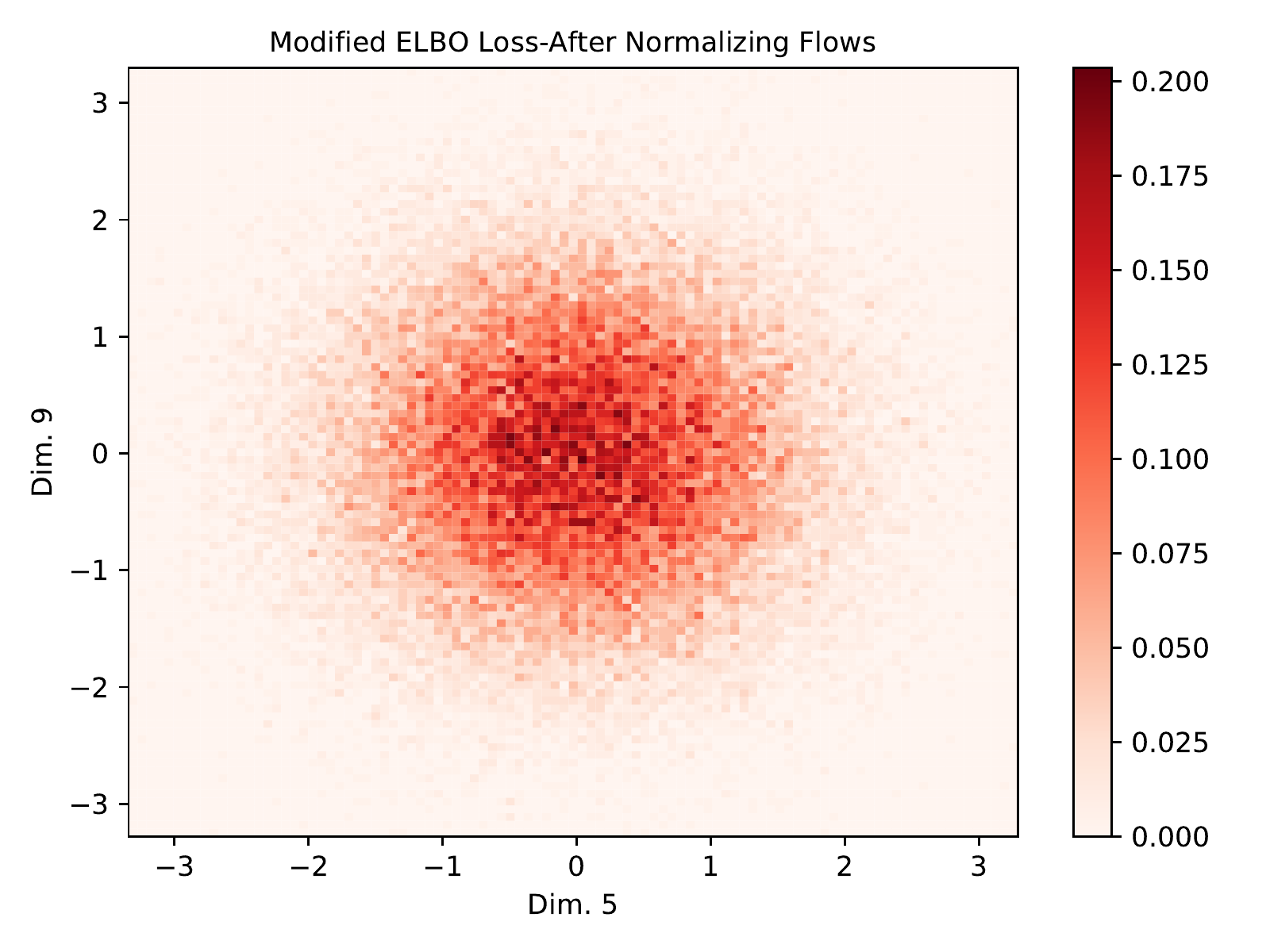}
    \caption{Latent space visualization by making histograms across arbitrarily chosen dimensions $5$ and $9$, before (left) and after normalizing flow transformations (right) with our loss function (top) and the modified ELBO loss (bottom).
    \label{fig:latentSpaceHist}}
\end{figure}

\begin{figure}[ht!]
    \centering
    \includegraphics[width=0.38\linewidth]{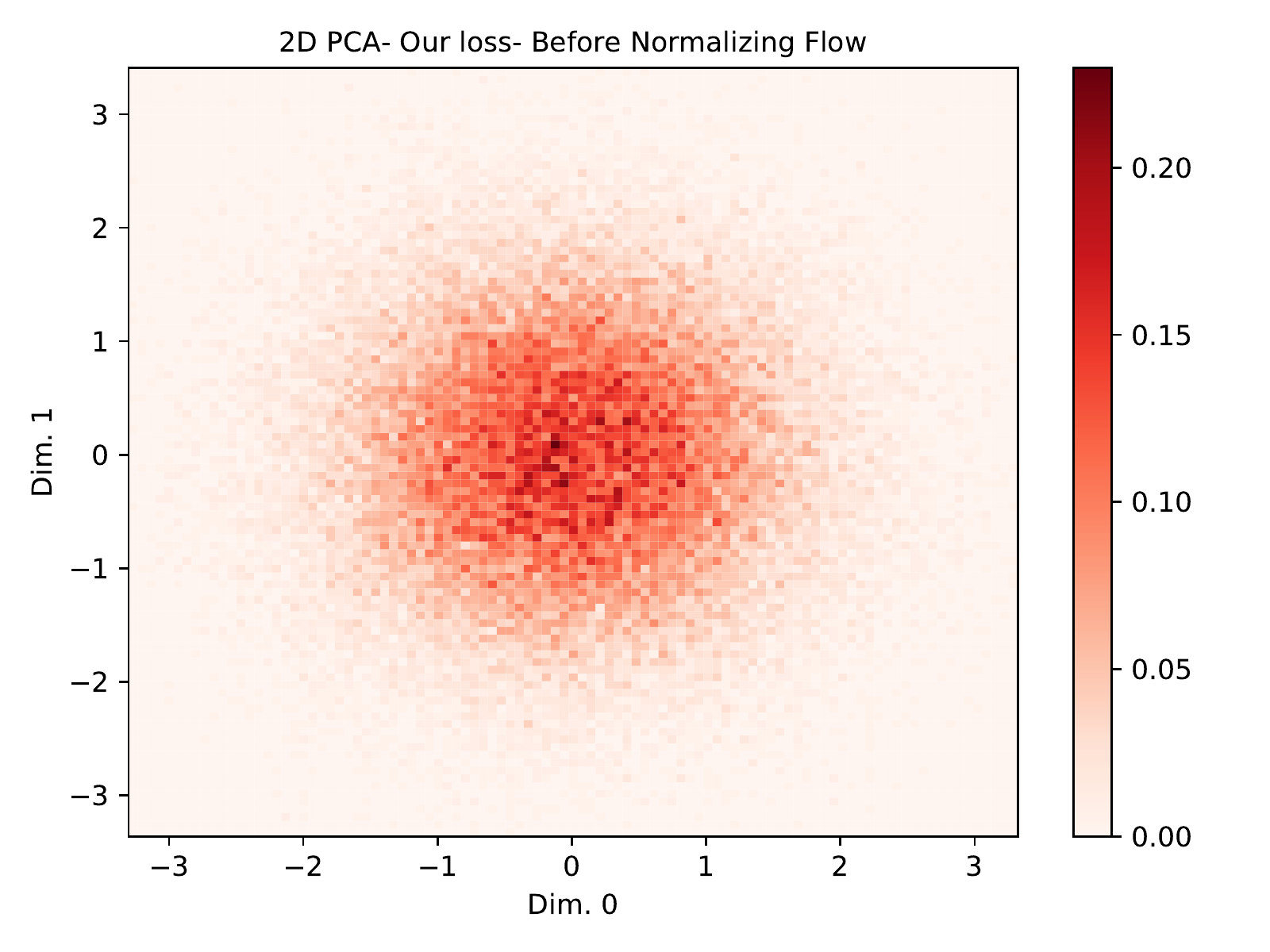}
    \includegraphics[width=0.38\linewidth]{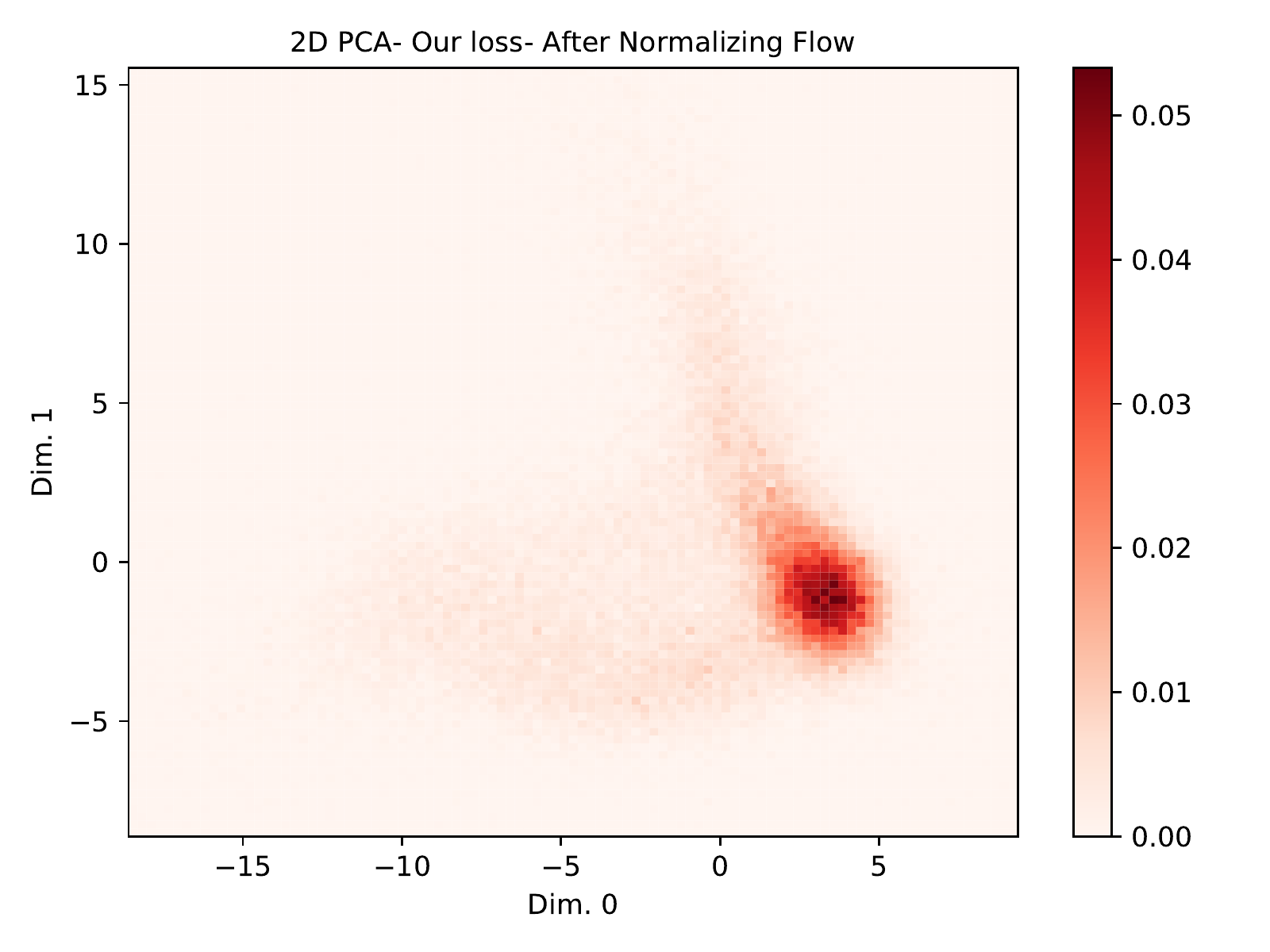}\\
    \includegraphics[width=0.38\linewidth]{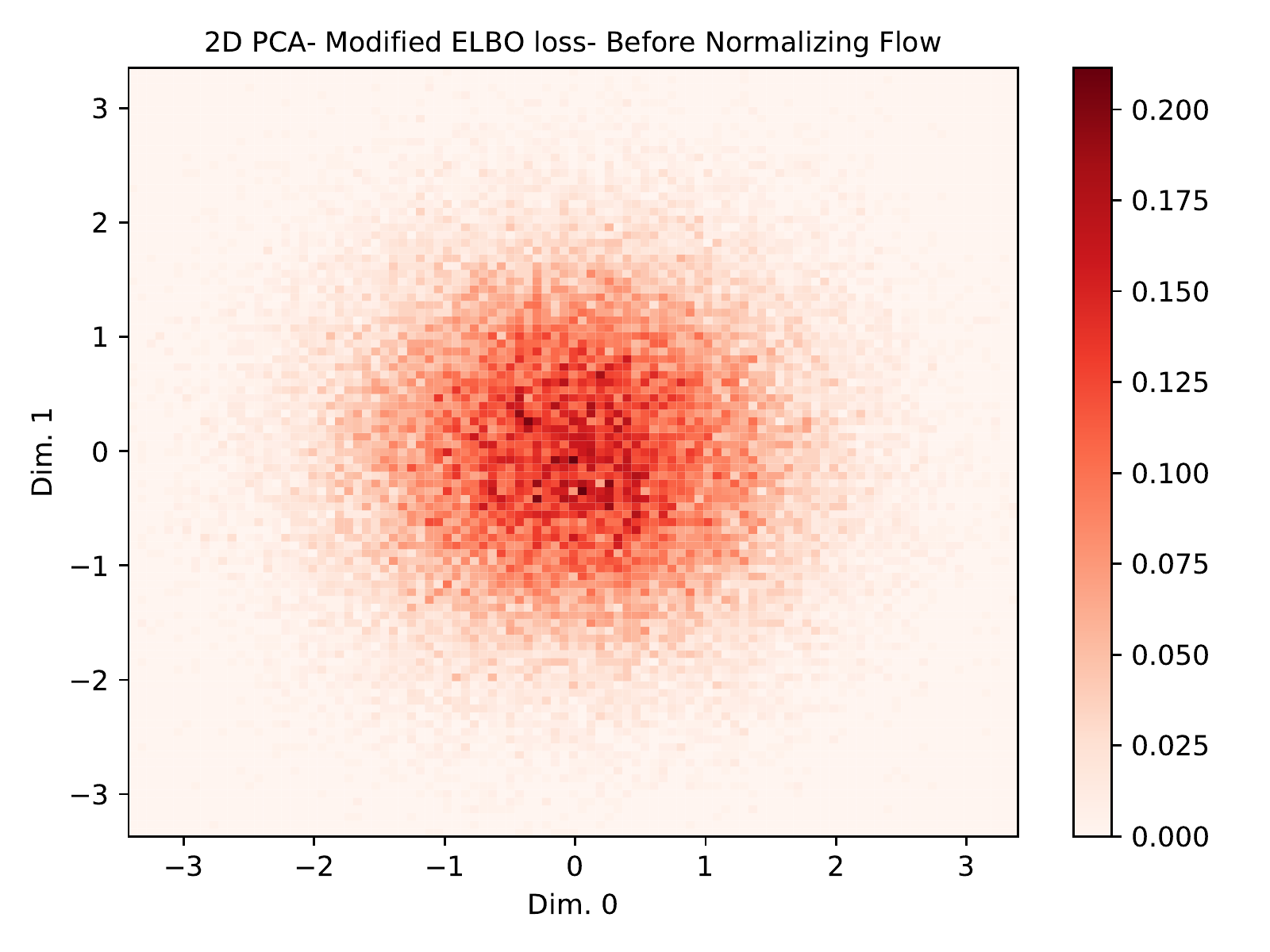}
    \includegraphics[width=0.38\linewidth]{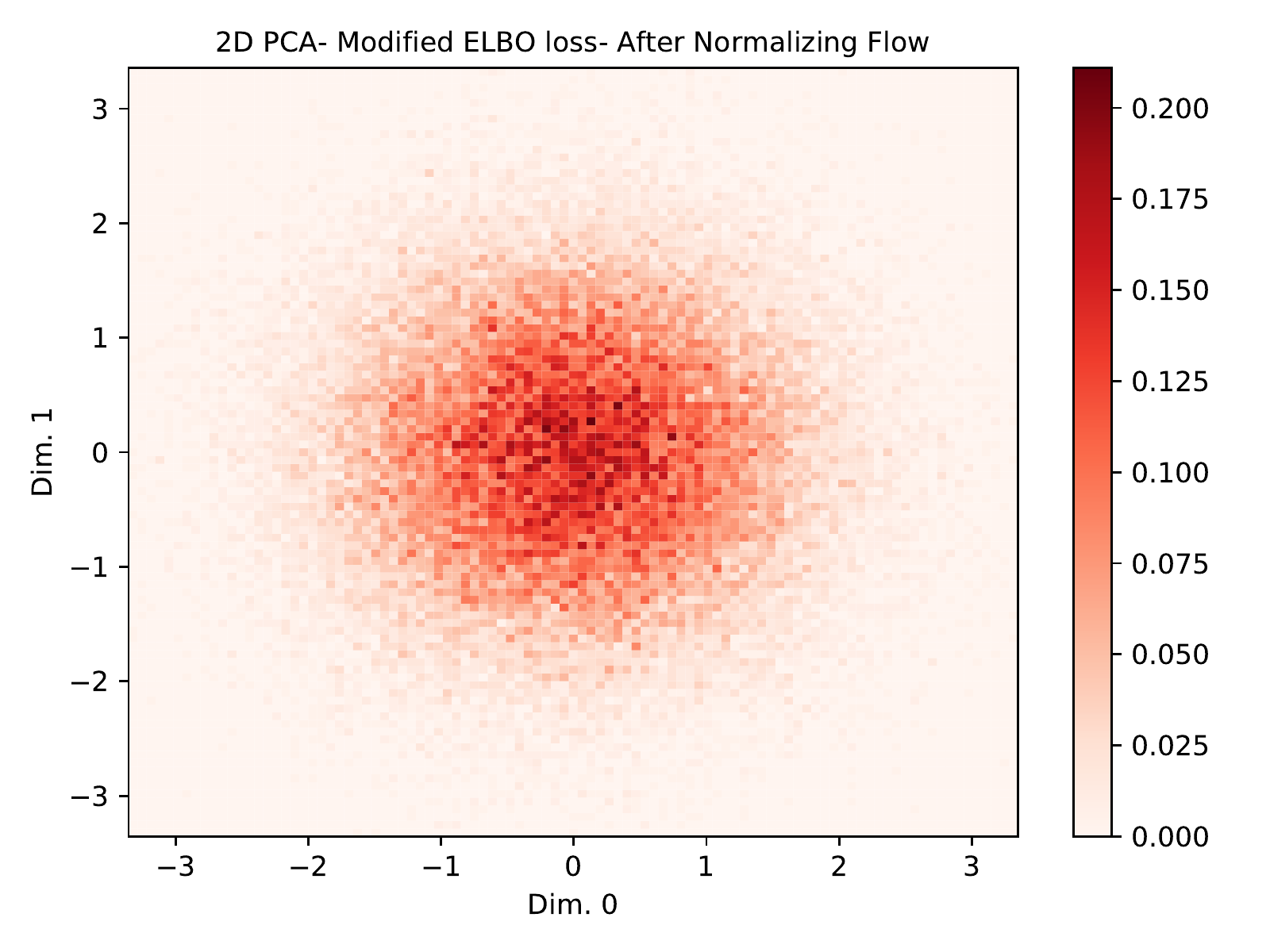}
    \caption{Latent space visualization after $2$D PCA, before (left) and after normalizing flow transformations (right) with our loss function (top) and the modified ELBO loss (bottom).
    \label{fig:latentSpacePCA}}
\end{figure}

We also show a comparison of the latent space obtained from the trainings with the two different loss functions, as described in the previous section of supplementary material. 
We see that our loss function results in a more complex, non-Gaussian distribution compared to the modified ELBO loss, and this is desirable to improve anomaly detection performance using VAEs. 
It is important to note that using our loss function may not necessarily correspond to better reconstruction of the input for the trained class (background samples) or better generation, but it rather contributes to a larger separation between the trained class and the non-trained class (signal samples) by making it harder for the decoder to reconstruct signal samples. 
As a result the anomaly identification performance increases regardless of whether the reconstruction of the background samples improves or not. 

\fi
\end{document}